\documentclass[11pt, a4paper, oneside]{article}

\pdfoutput=1

\usepackage[english]{babel}
\usepackage[utf8]{inputenc}
\usepackage{amsmath}
\usepackage{amssymb}
\usepackage{amsthm}
\usepackage{cite}
\usepackage{xcolor}
\usepackage[ruled,vlined]{algorithm2e}
\usepackage{float}
\usepackage{booktabs}
\usepackage{graphicx}
\usepackage{colortbl}
\usepackage[hidelinks]{hyperref}
\usepackage{adjustbox}
\usepackage{geometry}

\usepackage{geometry}

\usepackage{natbib}
\SetKw{And}{and}
\SetKw{Break}{break}

\usepackage{authblk}
\title{Resampling-Based Multisplit Inference for High-Dimensional Regression}
\author[1]{Anna Vesely\footnote{\href{mailto:anna.vesely@unipd.it}{anna.vesely@unipd.it}}}
\author[2]{Jelle J.~Goeman\footnote{\href{mailto:j.j.goeman@lumc.nl}{j.j.goeman@lumc.nl}}}
\author[1]{ Livio Finos\footnote{\href{mailto:livio.finos@unipd.it}{livio.finos@unipd.it}}}
\affil[1]{\footnotesize Department of Developmental Psychology and Socialization, University of Padova}
\affil[2]{\footnotesize Department of Biomedical Data Sciences, Leiden University Medical Center}
\date{}                     
\date{}

\newtheorem{theorem}{Theorem}

\newtheorem{lemma}{Lemma}
\newtheorem{proposition}{Proposition}
\newtheorem{assumption}{Assumption}

\newcommand{\Rset}{\mathbb{R}}
\newcommand{\mylim}{n\to\infty}
\newcommand{\limdistr}[1][n\to\infty]{\xrightarrow[\mylim]{\text{d}}}

\newcommand{\myexp}{\mathbb{E}} 
\newcommand{\myvar}{\text{var}} 
\newcommand{\mycov}{\text{cov}} 

\newcommand{\bevec}{\boldsymbol{\beta}}
\newcommand{\epsvec}{\boldsymbol{\varepsilon}}
\newcommand{\yvec}{\mathbf{Y}}
\newcommand{\xmat}{\mathbf{X}}
\newcommand{\ve}[1][j]{\mathbf{X}_{#1}} 
\newcommand{\ze}[1][j]{\mathbf{X}_{-#1}} 
\newcommand{\re}[1][j]{\mathbf{R}_{-#1}} 
\newcommand{\he}[1][j]{\mathbf{H}_{-#1}} 

\newcommand{\fmat}[1][b]{\mathbf{F}_{#1}}

\newcommand{\Te}[1][j]{T_{#1}} 
\newcommand{\tjo}[1][b]{\mathbf{t}_{j#1}}
\newcommand{\Tevect}[1][j]{\mathbf{T}_{#1}}

\newcommand{\Ts}[1][j]{\tilde{T}_{#1}} 
\newcommand{\tj}[1][b]{\mathbf{\tilde{t}}_{j#1}}
\newcommand{\Tsvect}[1][j]{\tilde{\mathbf{T}}_{#1}}

\newcommand{\Ue}[1][j]{U_{#1}} 
\newcommand{\ujo}[1][b]{\mathbf{u}_{j#1}}
\newcommand{\Uevect}[1][j]{\mathbf{U}_{#1}}

\newcommand{\Us}[1][j]{\tilde{U}_{#1}} 

\newcommand{\Ve}[1][j]{V_{#1}} 
\newcommand{\vjo}[1][b]{\mathbf{v}_{j#1}}
\newcommand{\Vevect}[1][j]{\mathbf{V}_{#1}}
\newcommand{\reb}[1][j]{\mathbf{\bar{R}}_{-#1}} 

\newcommand{\Vs}[1][j]{\tilde{V}_{#1}} 

\newcommand{\multinormal}[1][n]{\mathcal{N}_{#1}}
\newcommand{\matrixnormal}[1][n]{\mathcal{MN}_{#1}}

\newcommand{\Mset}{\mathcal{M}}
\newcommand{\Sset}{\mathcal{S}}
\newcommand{\Nset}{\mathcal{M}_0}
\newcommand{\Aset}{\mathcal{A}}

\newcommand{\Din}[1][q]{\mathcal{D}^{#1}_0}
\newcommand{\Dout}[1][q]{\mathcal{D}^{#1}}

\begin{document}
\maketitle


\begin{abstract}
We propose a novel resampling-based method to construct an asymptotically exact test for any subset of hypotheses on coefficients in high-dimensional linear regression. It can be embedded into any multiple testing procedure to make confidence statements on relevant predictor variables.
The method constructs permutation test statistics for any individual hypothesis by means of repeated splits of the data and a variable selection technique; then it defines a test for any subset by suitably aggregating its variables' test statistics. The resulting procedure is extremely flexible, as it allows different selection techniques and several combining functions. We present it in two ways: an exact method and an approximate one, that requires less memory usage and shorter computation time, and can be scaled up to higher dimensions. We illustrate the performance of the method with simulations and the analysis of real gene expression data.
\end{abstract}

\smallskip
\noindent \textbf{Keywords:} high-dimensional linear regression, multiple testing, Multisplit, permutation test, score


\newpage
\section{Introduction}
In the framework of linear regression, interest usually lies in discovering relevant predictor variables and assessing statistical significance. However, many challenges arise in high-dimensional settings, where the number of variables is potentially much larger than the sample size. Different methods have been proposed in literature to obtain error control and significance \citep{singlesplit,multisplit,stabsel,buhlmann,zhang,hdpost,hdsiminf}; for a review, see \citet{hdreview}.

In this manuscript we present a novel procedure that provides an asymptotically exact test for any subset of hypotheses on coefficients in high-dimensional linear regression. We rely on the permutation framework, which has proven to be often more powerful than the parametric approach, especially when testing multiple hypotheses \citep{westyoung, pesarin, exact0, fdpbounds}. As the proposed procedure allows to test any subset of hypotheses, it can be embedded into multiple testing methods such as the maxT-method \citep{westyoung} and closed testing \citep{closed}. In particular, it allows a form of post-hoc inference if used within closed testing methods that give simultaneous confidence sets for the proportion of true discoveries (TDP) within subsets \citep{genovese2, exploratory, sumSome_paper}.

To construct our procedure, we employ the sample-splitting framework of the Multisplit \citep{multisplit}, a powerful method that exploits variable selection techniques to compute adjusted p-values for all variables' coefficients. The procedure repeatedly splits the data into two random subsets, using the first to select variables and the second to obtain raw p-values via ordinary least squares (OLS) estimation; then the raw p-values are adjusted and aggregated over the splits. Moreover, we rely on the permutation test introduced by \citet{score} and \citet{finosflip} for parameters in generalized linear models (GLMs), based on sign-flipping score contributions. The test is asymptotically exact, and allows for estimation of other unknown parameters. Moreover, it is robust even for some misspecifications of the model such as overdispersion and heteroscedasticity and, in some cases, in presence of ignored nuisance parameters. Under the correct model, the power has been shown to be comparable to the parametric counterpart.

First, we introduce an approach similar to \citet{multisplit} that constructs permutation test statistics for each individual variable by means of repeated splits of the data, variable selection, and the test of \citet{score} and \citet{finosflip}. Then we define an asymptotically exact test for any subset by efficiently aggregating the individual variables' statistics with a suitable function; different combining functions are possible, including the maximum and weighted sums. Finally we propose an approximation of the procedure that requires less memory usage and shorter computation time, and can be scaled up to higher dimensions.

The structure of the manuscript is as follows. In Section \ref{hdlr} we introduce the model and its assumptions, as well as the Multisplit method \citep{multisplit} and the test based on sign-flipping score contributions \citep{score,finosflip}. Then we define the method and the approximate version in Sections \ref{rbmulti} and \ref{rbmultia}, respectively. Finally, in Sections \ref{multisims} ans \ref{multiribo} we explore the behavior of the proposed methods on simulated and real data. Proofs, algorithmic implementation and some additional results are postponed to the appendix.


\section{High-dimensional linear regression}\label{hdlr}
In this section we introduce notation and assumptions, as well as the two building blocks of the proposed procedure: the Multisplit method \citep{multisplit} and the permutation test based on sign-flipping score contributions \citep{score,finosflip}. Throughout the paper, we will refer to null hypotheses simply as hypotheses. Moreover, we will denote with capital letters hypotheses, variables and some numerical constants, leaving the distinction to context.

We consider a linear regression framework with $n$ observations and $m$ variables, potentially high-dimensional ($n \ll m$). The model is
\begin{align}
\yvec=\xmat\bevec + \epsvec,\qquad\epsvec\sim\multinormal(\mathbf{0},\sigma^2 \mathbf{I}) \label{def:model}
\end{align}
where $\yvec\in\Rset^n$ is the response vector, $\xmat\in\Rset^{n\times m}$ is a fixed design matrix, $\bevec\in\Rset^m$ is the vector of unknown coefficients and $\epsvec\in\Rset^n$ is a random error vector. Here $\multinormal$ denotes the multivariate normal distribution of size $n$, and $\mathbf{I}$ is the identity matrix. Even though some results will be asymptotic in $n$, for simplicity of notation we omit any dependence on it. Moreover, we assume that $\xmat^\top \xmat /n$ converges to a finite positive semi-definite matrix as $\mylim$.

We are interested in exploring which variables in $\xmat$ are active, meaning that they have non-null coefficients and so an effect on the response $\yvec$. Let $\Mset=\{1,\ldots,m\}$ be the set of variable indices, and $\Nset=\{j\in\Mset\,:\,\beta_j = 0\}$ the unknown subset corresponding to inactive variables. For any $j\in\Mset$, we may define the hypothesis
\begin{align*}
H_j\,:\,\beta_j = 0,
\end{align*}
that is true when $j\in\Nset$, regardless of the value of other variables' coefficients. We want to study more variables taken together, i.e., test intersection hypotheses of the form
\begin{align}\label{def:intershp}
H_\Sset=\bigcap_{j\in\Sset}H_j\,:\,\beta_j=0\text{ for all } j\in\Sset,\qquad \Sset\subseteq\Mset,\;\Sset\neq\emptyset
\end{align}
with significance level $\alpha\in[0,1)$. $H_\Sset$ is true if all variables in $\Sset$ are inactive, i.e., $\Sset\subseteq\Nset$.

To construct a test for any $H_\Sset$, we will rely on a selection procedure that estimates the set of active variables $\Mset\setminus\Nset$, returning a subset $\Aset\subseteq\Mset$. As in \citet{multisplit}, we assume that this procedure has the following properties.

\begin{assumption}[sparsity]\label{A:sparsity}
The number of selected variables is at most half the sample size: $|\Aset|\leq n/2$, where $|\cdot|$ denotes the size of a set.
\end{assumption}

\begin{assumption}[screening property]\label{A:screening}
Asymptotically, all truly active variables are selected:
\[\lim_{\mylim} P(\Mset\setminus\Nset \subseteq \Aset)=1.\]
\end{assumption}

The ideal selection procedure, for which the screening property always holds, is an oracle method that always selects all truly active variables, plus eventually some others. Even though such a procedure is not available in practice, we will use it in simulations to show the performance of the proposed method when Assumptions \ref{A:sparsity} and \ref{A:screening} are ensured. When studying real data, we suggest using the Lasso \citep{lassotib} with a suitable calibration of the $\lambda$ parameter, so that it selects enough variables for the screening property to be likely. If $m_1$ is an estimate of the expected number of active variables, we recommend choosing $\lambda$ so that the Lasso selects $\min(2m_1,n/2)$ variables. If there is no information available to give an estimate $m_1$, we recommend selecting as many variables as possible, i.e., $n/2$.


\subsection{Multisplit}\label{descrmultisplit}
The Multisplit method of \citet{multisplit} is a multiple testing procedure for high-dimensional linear regression that provides adjusted p-values $p_j$ for each variable $j\in\Mset$. Building on a proposal of \citet{singlesplit}, the main idea is to repeatedly split the data into two subsets for a number $Q$ of times (e.g., the Authors use $Q=50$ in simulations). The first subset is used to perform variable selection, while the second is used to compute raw p-values for the selected variables. Then p-values are obtained by adjusting the raw p-values and aggregating over the $Q$ splits.

For each split $q\in\{1,\ldots,Q\}$, the $n$ observations are randomly partitioned into two subsets $\Din$ and $\Dout$ of equal size $n/2$. First, observations in $\Din$ are employed to obtain an estimate $\Aset^q\subseteq\Mset$ of the set of active variables, using a variable selection procedure for which Assumptions \ref{A:sparsity} and \ref{A:screening} are supposed to hold. Then observations in $\Dout$ are used to compute raw p-values $\tilde{p}^q_j$ for each $j\in\Aset^q$ via OLS estimation; raw p-values for non-selected variables are set to 1.

Finally, for each variable $j\in\Mset$ the corresponding raw p-values are adjusted as
\begin{align}\label{def:adjunaggregated}
p_j^q=\min\{|\Aset^q|\,\tilde{p}_j^q,1\}\qquad (q\in\{1,\ldots,Q\})
\end{align}
and aggregated over the splits as
\begin{align}\label{def:adjaggregated}
p_j = \min\left\{1,\, (1-\log(\gamma_{\min})\inf_{\gamma\in (\gamma_{\min},1)} C_j(\gamma)\right\}
\end{align}
where
\[C_j(\gamma)=\min\left\{1,c_{\gamma}(p_j^1),\ldots,c_{\gamma}(p_j^Q)\right\},\]
$c_\gamma$ is the empirical $\gamma$-quantile function, and $\gamma_{\min}\in (0,1)$ is a lower bound for $\gamma$ (typically $\gamma_{\min}=0.05$).

The method identifies as active all variables $j$ with $p_j \leq \alpha$. \citet{multisplit} show that this procedure asymptotically controls the family-wise error rate (FWER) at level $\alpha$. Moreover, they prove that the p-values can be used to define a procedure with asymptotic control of the false discovery rate (FDR), extending the methodology of \citet{benjamini}.

An algorithm for the Multisplit method is presented in Section \ref{alg_multi}.


\subsection{Sign-flipping score contributions}\label{sfsc}
In this section we introduce the permutation test proposed by \citet{score} and \citet{finosflip} restricting to our setting, i.e., linear regression. Throughout this section suppose that the framework is low-dimensional with $n>m$.

Fix any variable $j\in\Mset$. To test the individual hypothesis $H_j$, \citet{score} provide a permutation test constructed from the absolute value of the effective score
\[\Te^1=\tjo[1]^\top\yvec,\qquad \tjo[1]=\frac{1}{\sqrt{n}}\re\ve\in\Rset^n,\]
where $\ve\in\Rset^n$ and $\ze\in\Rset^{n\times (m-1)}$ are obtained from the design matrix $\xmat$ by taking and removing the $j$-th column, respectively, and
\begin{align}\label{def:rmmatrix}
\re=\mathbf{I} -\ze(\ze^\top\ze)^{-1}\ze^\top\in\Rset^{n\times n}
\end{align}
is the residual maker matrix defined from $\ze$.

A critical value for the test statistic $|\Te^1|$ is constructed using $B$ random transformations of the data. The value of $B$ does not need to grow with $m$; larger values tend to give more power, but to have non-zero power we only need $B\geq 1/\alpha$. Hence let $\fmat[1],\ldots,\fmat[B]\in\Rset^{n\times n}$ be diagonal sign-flipping matrices, where $\fmat[1]=\mathbf{I}$ is the identity, while the diagonal elements of the other matrices are independently and uniformly drawn from $\{-1,1\}$. These matrices define
\begin{align}\label{def:Te}
\Te^b = \tjo^\top\yvec,\qquad\tjo =\frac{1}{\sqrt{n}}\re \fmat \re \ve \in\Rset^n\qquad (b\in\{1,\ldots,B\}).
\end{align}
Then a critical value is $|\Te^{(\lceil (1-\alpha)B\rceil)}|$, where $|\Te^{(1)}|\leq\ldots\leq|\Te^{(B)}|$ are the sorted values, and $\lceil\cdot\rceil$ denotes the ceiling function. As shown in the following proposition, the resulting test is asymptotically exact, but may be anti-conservative \citep{score,finosflip}.

\begin{proposition}\label{P:signflip}
The test that rejects $H_j$ when $|\Te^1| > |\Te^{(\lceil (1-\alpha)B\rceil)}|$ is asymptotically an $\alpha$-level test for any $j\in\Mset$. For finite $n$, it may be anti-conservative as
\[\myvar(\Te^1)\geq\myvar(\Te^b)\qquad (b\in\{1,\ldots,B\}).\]
\end{proposition}

From this framework, \citet{finosflip} construct a test that is exact for any sample size $n$. Observe that, for each transformation $b$,
\[\myvar(\Te^b\,|\,\fmat)=\sigma^2 \|\tjo\|^2\]
where $\sigma^2$ is unknown but common to all transformations, and $\|\tjo\|=(\tjo^\top\tjo)^{1/2}$ is the Euclidean norm of the known vector $\tjo$. An exact test can be obtained from the standardized scores
\begin{align}\label{def:Ts}
\Ts^b = \tj^\top\yvec,\qquad\tj=
\begin{cases}
\mathbf{0}\quad\text{if}\quad\|\tjo\|=0\\
\|\tjo\|^{-1}\tjo\qquad\text{otherwise.}
\end{cases}
\end{align}


\begin{theorem}\label{T:signflip2}
The test that rejects $H_j$ when $|\Ts^1| > |\Ts^{(\lceil (1-\alpha)B\rceil)}|$ is an $\alpha$-level test for any $j\in\Mset$.
\end{theorem}

In the next section we will rely on the ideas underlying Theorem \ref{T:signflip2}, as well as the Multisplit framework, to construct permutation test statistics for each variable in high-dimensional linear regression. Then we will show how these statistics may be employed to study sets of variables.


\section{Resampling-based inference}\label{rbmulti}

In this section we propose an asymptotically exact test for any intersection hypothesis $H_\Sset$, as given in \eqref{def:intershp}, valid even in high-dimensional settings. The method builds on the idea that we can efficiently construct a test statistic for $H_\Sset$ by combining statistics for the individual hypotheses $H_j$ with $j\in\Sset$ in a suitable way. We take as combining function any $g\,:\,\Rset^{|\Sset|}\longrightarrow\Rset$ which is increasing in each argument, such as the maximum or (weighted) sums.

First, we prove this in the low-dimensional setting of Section \ref{sfsc}. In this framework, \citet{score} provide a permutation test for any intersection hypothesis, but it requires to compute new test statistics for each set $\Sset$. In the following lemma we prove that an asymptotically exact test for $H_\Sset$ can be defined using a combination of the standardized scores
\begin{align}\label{def:combTs}
\Ts[\Sset]^b= g\left(|\Ts[j_1]^b|,\ldots, |\Ts[j_s]^b|\right)\qquad (\Sset =\{j_1,\ldots,j_s\},\,b\in\{1,\ldots,B\}).
\end{align}

\begin{lemma}\label{L:combinevars}
The test that rejects $H_\Sset$ when $\Ts[\Sset]^1 > \Ts[\Sset]^{(\lceil (1-\alpha)B\rceil)}$ is asymptotically an $\alpha$-level test for any non-empty $\Sset\subseteq\Mset$.
\end{lemma}

This way, to study all subsets it is sufficient to compute the $m$ individual statistics $\Ts[1],\ldots,\Ts[m]$ given in \eqref{def:Ts}. The lemma holds equivalently also for the non-standardized version in \eqref{def:Te}.

Now we provide an analogous method for the high-dimensional setting. Similarly to \citet{multisplit}, we rely on a variable selection procedure that is assumed to fulfill Assumptions \ref{A:sparsity} and \ref{A:screening}, and we split the data into two subsets for a number $Q$ of times. The first subset is used to select active variables, then the second subset is used to compute effective scores as in \eqref{def:Te}, sum these scores over the splits, and suitably standardize as in \eqref{def:Ts}.

Fix $B$ diagonal sign-flipping matrices $\fmat[1],\ldots,\fmat[B]\in\Rset^{n\times n}$, where $\fmat[1]=\mathbf{I}$, while the diagonal elements of the other matrices are independently and uniformly drawn from $\{-1,1\}$. For each split $q\in\{1,\ldots,Q\}$, we randomly divide observations into two equally-sized subsets $\Din$ and $\Dout$. First, we use observations in $\Din$ to estimate the set of active variables with $\Aset^q\subseteq\Mset$. Subsequently, we use observations in $\Dout$ and selected variables in $\Aset^q$ to compute effective scores for all transformations. Particular attention is to be paid to the use of transformations; if the same observation is in $\Dout[q_1]$ and $\Dout[q_2]$, then it must undergo the same sign-flipping transformations between the two splits. The procedure may be written as follows.

We restrict the design matrix $\xmat$ to observations in $\Dout$ and variables in $\Aset^q$, obtaining
\[\xmat^q = \xmat_{\Dout, \Aset^q}.\]
For any variable $j\in\Mset$, we define the split's residual maker matrix taking $\re^q=\mathbf{0}\in\Rset^{n\times n}$, and substituting
\begin{align}\label{def:splitresmat}
\mathbf{R}^q_{-j;\Dout,\Dout} =
\begin{cases}
\mathbf{I} - \ze^q (\ze^{q\top}\ze^q)^{-1} \ze^{q\top}\qquad\text{if}\qquad j\in \Aset^q\\
\mathbf{0}\qquad\text{otherwise.}
\end{cases}
\end{align}
This way, if $j$ is selected elements corresponding to observations in $\Dout$ are computed from the residual maker matrix defined from $\ze^q$ (see \eqref{def:rmmatrix}); otherwise we obtain the null matrix. The matrices $\re^q$ are used to construct the sum of the effective scores
\begin{align}
\Ue^b = \sum_{q=1}^Q\frac{1}{\sqrt{n}}\ve^\top\re^q \fmat\re^q \yvec=\ujo^\top \yvec,\qquad \ujo=\frac{1}{\sqrt{n}}\sum_{q=1}^Q\re^q \fmat\re^q\ve \label{def:Ue}
\end{align}
for any variable $j$ and any transformation $b$. Finally, we obtain a standardized score $\Us^b$ normalizing $\ujo$, as in \eqref{def:Ts}. Notice that $\Us^b=0$ if $j$ is never selected.

These individual test statistics $\Us^b$ can be combined to test any intersection hypothesis $H_\Sset$ analogously to Lemma \ref{L:combinevars}, using
\begin{align}\label{def:combTs_exact}
\Us[\Sset]^b= g\left(|\Us[j_1]^b|,\ldots,|\Us[j_s]^b|\right)\qquad (\Sset =\{j_1,\ldots,j_s\},\,b\in\{1,\ldots,B\}).
\end{align}

\begin{theorem}\label{T:multi_exact}
The test that rejects $H_\Sset$ when $\Us[\Sset]^1 > \Us[\Sset]^{(\lceil (1-\alpha)B\rceil)}$ is asymptotically an $\alpha$-level test for any non-empty $\Sset\subseteq\Mset$.
\end{theorem}

Similarly to the test of Theorem \ref{T:signflip2}, the test for any individual hypothesis $H_j$ is exact not only asymptotically, but for any sample size $n$.

To summarize, we have proposed a method to construct permutation test statistics for all variables in high-dimensional linear regression, using $Q$ random splits and $B$ random transformations. These individual test statistics are sufficient to define an asymptotically exact permutation test for any intersection hypothesis $H_\Sset$. Indeed, by Theorem \ref{T:multi_exact} such a test can be obtained combining the statistics for the variables in $\Sset$ through any function $g$ that is increasing in each argument. As the method provides a test for all $H_\Sset$, it can be embedded into multiple testing methods such as the maxT-method \citep{westyoung} (if $g=\max$) and closed testing \citep{closed}. In particular, it can be used within closed testing procedures that give simultaneous confidence sets for the TDP such as \citet{exploratory} and, if $g$ is a sum, \citet{sumSome_paper}.


In Section \ref{alg_splitflipe} we provide an algorithm for the method. In the worst case, it requires a number of operations of order $n^4QB$, and memory usage of order $n^2Q$. The high memory usage is due to the need to store, for any variable $j$, the residual maker matrices $\re^1,\ldots,\re^Q$. In the following section we will introduce a new procedure that is less computationally expensive.



\section{Approximate method}\label{rbmultia}
Section \ref{rbmulti} provides a procedure to test any intersection hypothesis $H_\Sset$ in high-dimensional linear regression. As the method requires intensive memory usage, in this section we propose an approximation that is less expensive. We prove that the resulting approximate method defines an asymptotically exact test, then in the next sections we will study the performance of the two methods, exact and approximate, through simulations and the analysis of real data.

The new approximate method relies on the same procedure introduced in the previous section, but defines new test statistics for which we no longer need to save all the residual maker matrices $\re^1,\ldots,\re^Q$. For any variable $j\in\Mset$ and each transformation $b\in\{1,\ldots,B\}$, instead of summing the splits' effective scores, we first sum the residual maker matrices
\begin{align}\label{sumr}
\reb=\sum_{q=1}^Q \re^{q}
\end{align}
then use the resulting matrix to compute an overall score
\begin{align}
\Ve^b =\vjo^\top\yvec,\qquad \vjo=\frac{1}{\sqrt{n}}\reb \fmat\reb\ve\label{def:Ve}.
\end{align}
Finally, we construct a standardized score $\Vs^b$ normalizing $\vjo$ as in \eqref{def:Ts}.

As in the previous section, the individual test statistics $\Vs^b$ can be combined to test any intersection hypothesis $H_\Sset$, using
\begin{align}\label{def:combTs_appr}
\Vs[\Sset]^b= g\left(|\Vs[j_1]^b|,\ldots,|\Vs[j_s]^b|\right)\qquad (\Sset =\{j_1,\ldots,j_s\},\,b\in\{1,\ldots,B\})
\end{align}
where $g\,:\,\Rset^{|\Sset|}\longrightarrow\Rset$ is a function increasing in each argument.

\begin{theorem}\label{T:multi_appr}
The test that rejects $H_\Sset$ when $\Vs[\Sset]^1 > \Vs[\Sset]^{(\omega)}$ is asymptotically an $\alpha$-level test for any non-empty $\Sset\subseteq\Mset$.
\end{theorem}

In conclusion, Theorem \ref{T:multi_exact} gives an asymptotically valid, but computationally intensive, procedure to test intersection hypotheses; Theorem \ref{T:multi_appr} provides a less expensive procedure based on an approximation. In both methods, computing the test statistics for the individual hypotheses $H_1,\ldots,H_m$ is sufficient to test any intersection hypothesis $H_\Sset$. As a consequence, the methods can be used within multiple testing procedures, as observed in Section \ref{rbmulti}. In the following sections we will use simulated and real data to study the behavior of the approximate method, comparing it to the exact method and the Multisplit of \citet{multisplit}, as well as investigating the role of the variable selection procedure. We show in particular that the error control holds in most of the considered settings, even with finite sample size.

An algorithm for the approximate method is provided in Section \ref{alg_splitflipa}. The computational complexity is lower than the exact method, but still polynomial in $n$ and linear both in $Q$ and in $B$; the memory usage is of order $n^2$. As memory operations (write and read) affect the running time of an algorithm, the approximate method will prove to be much faster than the exact.


\section{Simulations}\label{multisims}
We use simulations to explore the performance of the proposed exact and approximate methods of Sections \ref{rbmulti} and \ref{rbmultia}. First we compare the two methods, then we further investigate the behavior of the approximate method, comparing it to the Multisplit \citep{multisplit} and using different variable selection procedures. We correct for multiplicity with the maxT-method \citep{westyoung}, corresponding to the combining function $g=\max$ in \eqref{def:combTs_exact} and \eqref{def:combTs_appr}. The proposed methods and the Multisplit are implemented in the packages \texttt{splitFlip} \citep{splitFlip} and \texttt{hdi} \citep{hdipackage} developed in R \citep{rsoftware}, respectively.

We use the same simulation settings proposed in \citet{multisplit}. The design matrix $\xmat$ is defined in two ways. First we take the real $71\times 4{,}088$ design matrix of the \texttt{riboflavin} dataset from the \texttt{R} package \texttt{hdi} \citep{hdipackage}, which contains gene expression levels of \textit{Bacillus subtilis}. Then we simulate a $n\times m$ Toeplitz design matrix coming from a centered multivariate normal distribution with $\mycov(\xmat_j,\xmat_h)=\rho^{|j-h|}$ for $j,h\in\Mset$. Subsequently, the response variable is computed as in \eqref{def:model}, where the coefficient vector $\bevec$ is such that $m_1$ elements are non-null, with values either all equal to 1 (uniform-strength setting) or equal to $1,2,\ldots,m_1$ (increasing-strength setting). The error standard deviation $\sigma$ is computed so that the signal-to-noise ratio is SNR.

We analyze the set $\Mset$ of all variables with significance level $\alpha$, using $B$ random sign-flipping transformations and $Q$ random splits of the data. We consider two variable selection procedures that select $2m_1$ variables: an oracle method where all truly active variables are always selected, and the Lasso with suitable $\lambda$-calibration. As observed in Section \ref{hdlr}, the oracle selection allows to emphasize the behavior of the proposed methods when assumptions are met, while the Lasso allows to appreciate the performance in practical applications, when oracle selection is not feasible.

As a basic scenario, we fix $m_1=5$, SNR $=4$ and $Q=50$, and we use the oracle selection procedure; moreover, we take $n=100$, $m=100$ and $\rho\in\{0,0.2,0.5,0.7,0.9\}$ for the simulated design matrix. Then we expand this scenario, varying some of the parameters in the different analyses. For each setting, we simulate data $1000$ times. We compute the number of rejections as the mean over the simulations, and the FWER as the proportion of simulations where the method rejects $H_j$ for at least one inactive variable $j$. Results are shown only for the uniform-strength setting, as those for the increasing-strength setting display the same behavior.


\subsection{Approximate and exact}\label{multisims1}
We compare the approximate method of Section \ref{rbmultia} with the exact method of Section \ref{rbmulti}, expanding the basic scenario with $Q\in\{10,50\}$. Results for the real design matrix are in Table \ref{Table:EA_tot}. In this case, both methods control the FWER; the approximate is slightly less powerful, but faster.

Results for the simulated design matrix are shown in Appendix \ref{appen:simmulti1}. The FWER is always controlled by the exact method, while it is slightly higher for the approximate in one setting ($\rho=0.7$ and $Q=50$). In terms of number of rejections, the approximate method is close to the exact, especially when the number of splits is high. The computation time is at most 32 seconds for the approximate, and 168 seconds for the exact.

\begin{table}
\centering
\caption{Real design matrix: results for the approximate and exact methods using $Q$ splits.}
\label{Table:EA_tot}
\begin{tabular}{ccccc}
\toprule
  & \multicolumn{2}{c}{approximate} & \multicolumn{2}{c}{exact}\\
$Q$  & 10 & 50 & 10 & 50\\
\midrule
FWER  & 0.046 & 0.048 & 0.045 & 0.048\\
rejections  & 3.1 & 3.4 & 3.9 & 3.9\\
time (s)  & 7.4 & 32.3 & 12.8 & 59.8\\
\bottomrule
\end{tabular}
\end{table}


\subsection{Approximate and Multisplit}\label{multisims2}
Now we compare the approximate method with the Multisplit of \citet{multisplit}. We expand the basic scenario taking $m_1\in\{5,10\}$ and $\text{SNR}\in\{0.25,1,4,16\}$, as well as $m\in\{100,1000\}$ for the simulated design matrix. The settings with $\rho=0.5$ correspond to those investigated in \citet{multisplit}.

\begin{figure}
\centering
\includegraphics[width=\textwidth]{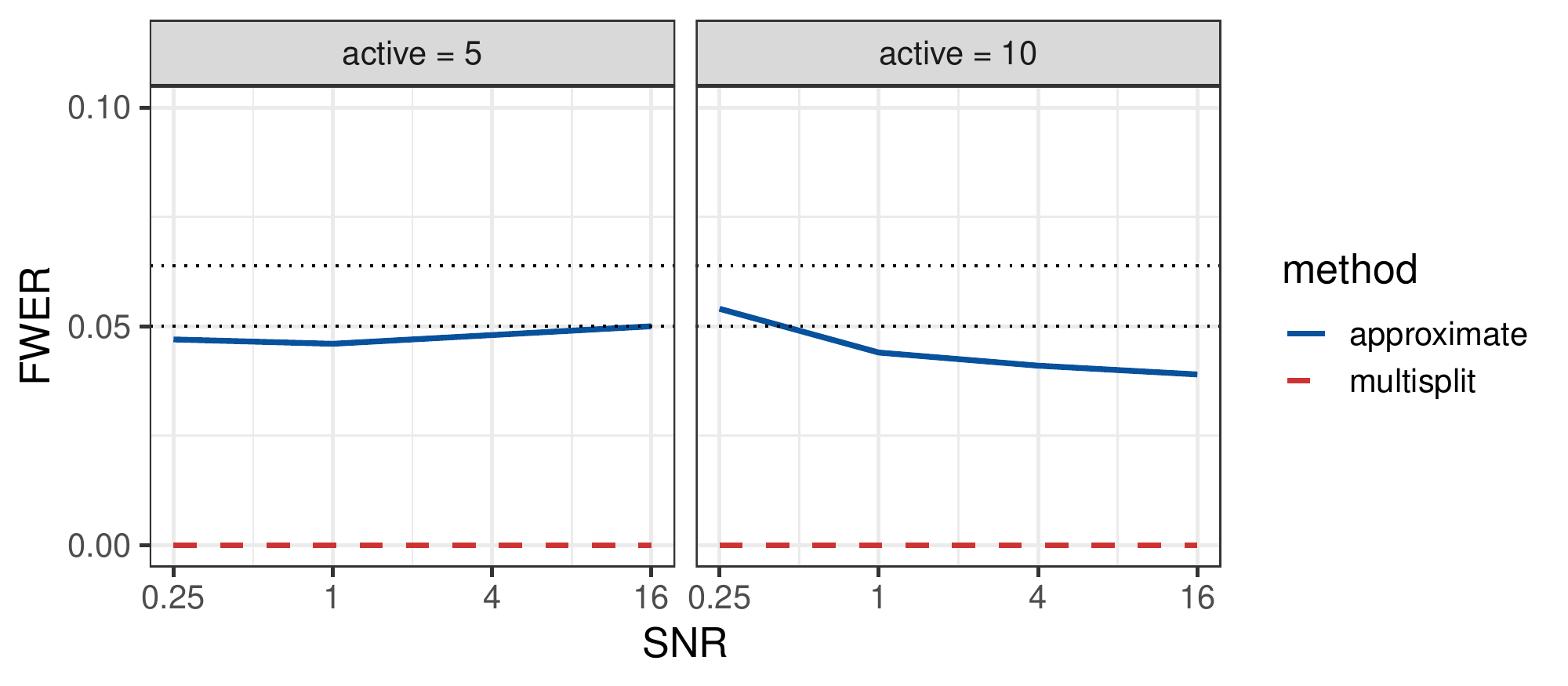}
\caption{Real design matrix: FWER by signal-to-noise ratio SNR (log scale), for the approximate method and the Multisplit. \textit{Active} denotes the number of active variables. The dotted lines correspond to the significance level $\alpha=0.05$ and an upper bound ($\alpha$ plus two standard deviations, approximately 0.063).}
\label{Plot:MAfwerreal}
\end{figure}

\begin{figure}
\centering
\includegraphics[width=\textwidth]{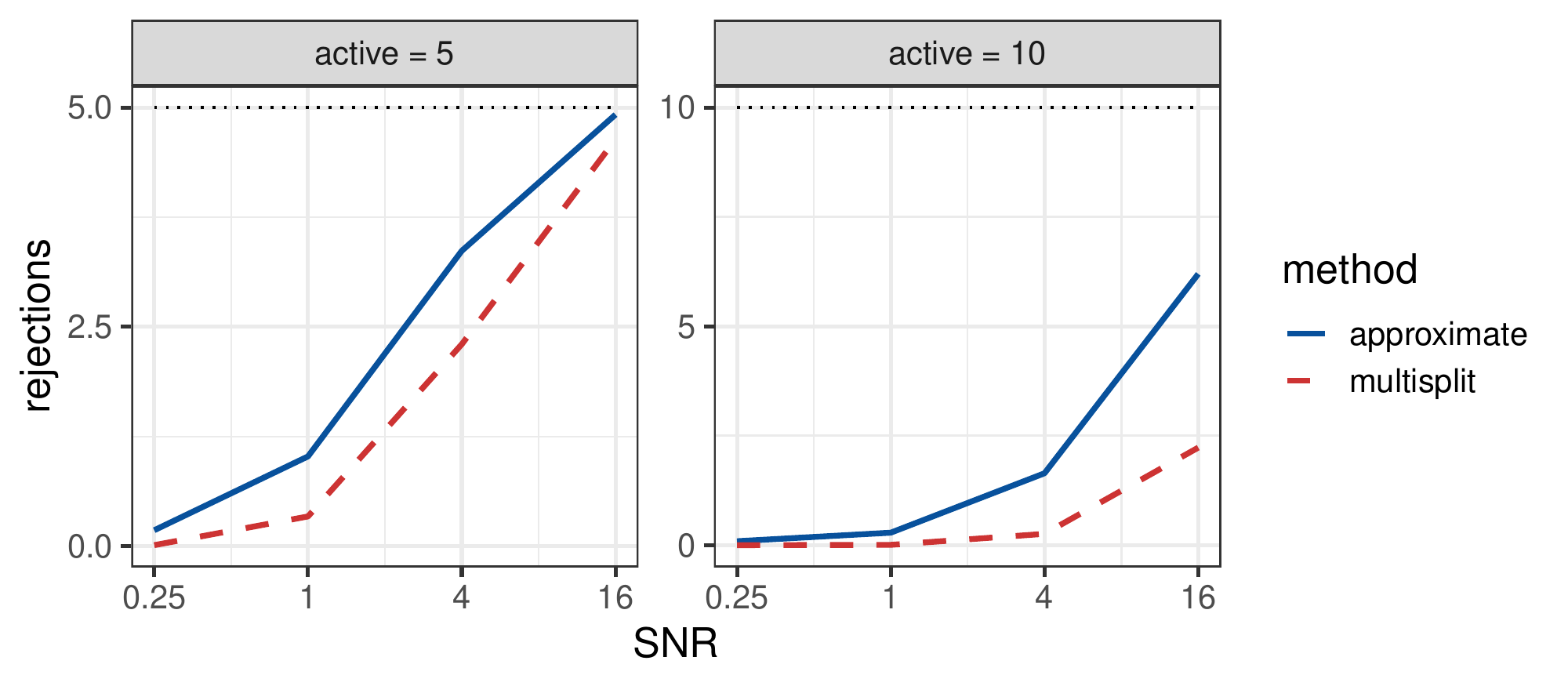}
\caption{Real design matrix: number of rejections by signal-to-noise ratio SNR (log scale), for the approximate method and the Multisplit. \textit{Active} denotes the number of active variables. The dotted line corresponds to \textit{active}.}
\label{Plot:MAtdpreal}
\end{figure}

Figures \ref{Plot:MAtdpreal} and \ref{Plot:MAfwerreal} contain results for the real design matrix, for which the approximate method always controls the FWER and is more powerful than the Multisplit.

Considering the simulated design matrix, for which results are shown in Appendix \ref{appen:simmulti2}, the proposed method control the FWER when the covariance parameter $\rho$ is not too high, or the signal-to-noise ratio SNR is particularly low. Among the scenarios where the FWER is controlled, the method is always more powerful than the Multisplit, with greatest differences when $\rho$ and SNR are low.

Computation times, shown in Appendix \ref{appen:simmulti2}, are feasible, never exceeding 3 minutes.

\subsection{Oracle and Lasso}\label{multisims3}
Finally, we examine the approximate method for different sample sizes $n$, using both selection procedures illustrated in Section \ref{hdlr}: oracle and Lasso \citep{lassotib}. The oracle is defined so that it always selects the $m_1$ truly active variables, plus $m_1$ others chosen at random. The Lasso selects the same number $2m_1$ of variables, with suitable $\lambda$-calibration.

\begin{figure}
\centering
\includegraphics[width=\textwidth]{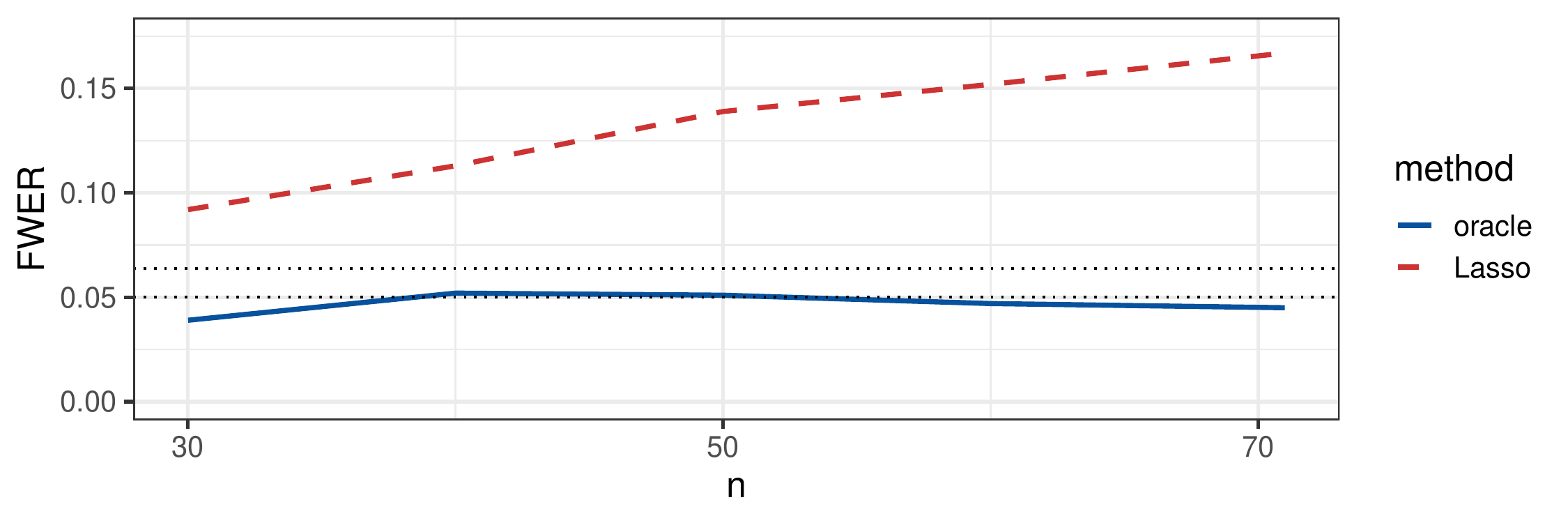}
\caption{Real design matrix: FWER by sample size $n$, for the approximate method using oracle selection and Lasso. The dotted lines correspond to the significance level $\alpha=0.05$ and an upper bound ($\alpha$ plus two standard deviations, approximately 0.063).}
\label{Plot:LOfwerreal}
\end{figure}

For the real data, where only 71 observations are available, we take $n\in\{30,40,\ldots,70\}$. The oracle always controls the FWER, while the Lasso looses control for all sample sizes (Figure \ref{Plot:LOfwerreal}). As expected, the power of the oracle increases with $n$ (Figure \ref{Plot:LOtdpreal}). Similar results are obtained for the simulated design matrix (see Appendix \ref{appen:simmulti3}) with $n\in\{30,40,\ldots,150\}$, where the FWER is always controlled by the oracle but not by the Lasso, which fails in some scenarios.

These results underline that particular attention must be paid to the choice of the selection method. Indeed, the asymptotic error control of the proposed methods (Theorems \ref{T:multi_exact} and \ref{T:multi_appr}) relies on the screening property given in Assumption \ref{A:screening}, and is not ensured when the property is not fulfilled.

\begin{figure}
\centering
\includegraphics[width=\textwidth]{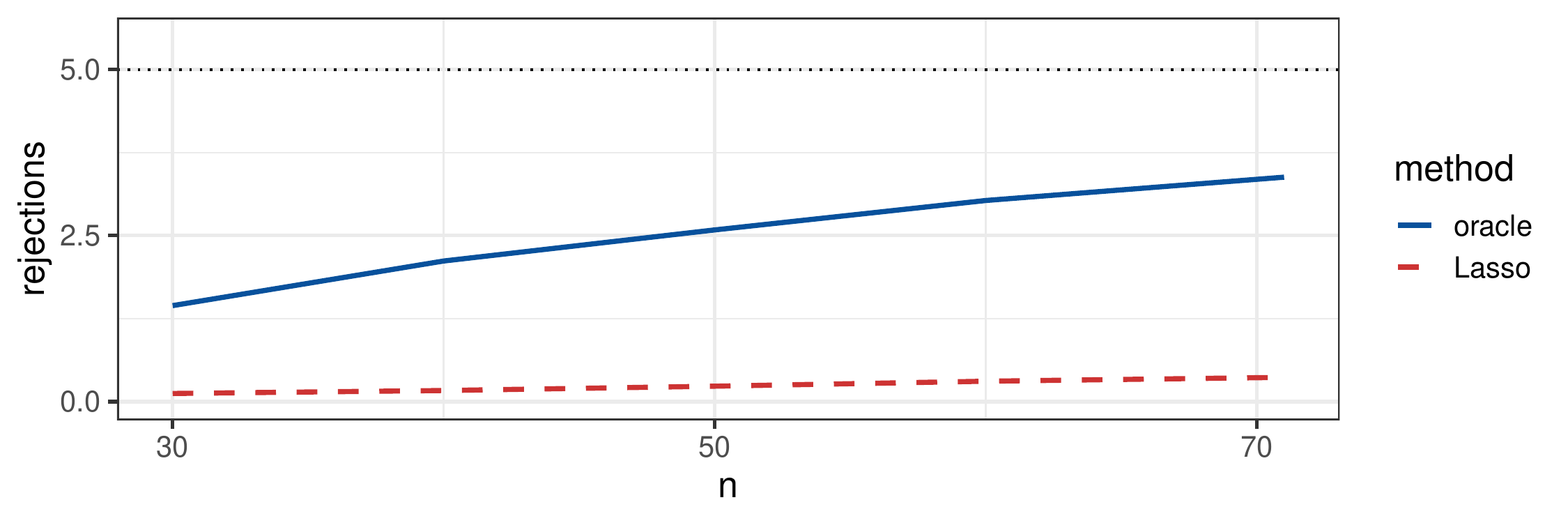}
\caption{Real design matrix: number of rejections by sample size $n$, for the approximate method using oracle selection and Lasso. The dotted line corresponds to the true number of active variables.}
\label{Plot:LOtdpreal}
\end{figure}


\section{Riboflavin data}\label{multiribo}
In this final section we study the performance of the approximate method of Section \ref{rbmultia} and the Multisplit \citep{multisplit} on real data. We analyze the \texttt{riboflavin} dataset from the \texttt{R} package \texttt{hdi} \citep{hdipackage}, containing data on riboflavin production by \textit{Bacillus subtilis}. Data consists of 71 observations of riboflavin production rate, as well as gene expression levels for $4{,}088$ genes. We assume a linear model where the first is the response and the latter are the predictors. We are interested in making inference on the influence of genes on the response, especially at the level of pathways, collections of genes associated with a specific biological process that interact with each other. We consider the 115 pathways contained in the KEGG database \citep{keggdb}.

We take $\alpha=0.05$, $Q=100$ and $B=200$. Moreover, we suppose that few genes influence the response, estimating this number with $m_1=5$. We perform four different analyses, all based on Lasso selection \citep{lassotib}: (a) Multisplit with 10-fold cross-validation (dafult of the package \texttt{hdi}); (b) Multisplit with the calibration of the $\lambda$ parameter suggested in Section \ref{hdlr}; (c) approximate method combined with the maxT-method \citep{westyoung}, as in the previous section ($g=\max$); (d) approximate method combined with the closed testing procedure of \citet{sumSome_paper}, for which the combining function $g$ in \eqref{def:combTs_appr} is the sum.

Analyses (a), (b) and (c) all give the same result, finding one single active gene: the negative regulatory protein YxlD. This gene is not contained in any of the considered pathways. Analysis (d) gives a lower ($1-\alpha$)-confidence bound for the number of true discoveries of 4 ($0.10\%$) among all genes. When studying the 115 pathways individually, however, we always obtain a lower confidence bound of zero. This may be due to the fact that active gene(s) do not appear in pathways, or to a signal too low to be detected; furthermore, the choice of the test statistic may be more suitable for denser signal \citep{sumSome_paper}.

The computation time needed to construct the test statistics for the approximate method is around 13 seconds on a standard PC. Then analysis (c) requires 3 additional seconds, and analysis (d) 30 seconds.


\section{Discussion}
We have considered the problem of testing multiple hypotheses in high-dimensional linear regression. Our proposed approach provides asymptotically valid resampling-based tests for any subset of hypotheses, which can be employed within multiple testing procedures to make confidence statements on active predictor variables.
For instance, it can be used within the maxT-method \citep{westyoung} and closed testing methods that give simultaneous confidence sets for the TDP of subsets \citep{genovese2, exploratory, sumSome_paper}.

To construct a test for a generic subset of hypotheses, we have provided a procedure that repeatedly splits the data into two random subsets, using the first to select variables, and the second to build permutation test statistics for each variable. Then statistics for any subset can be defined by aggregating individual statistics with different functions, including the maximum and weighted sums. The computational complexity is linear in the number of splits and permutations, and polynomial in the sample size in the worst case. As the method has intensive memory usage, requiring to store many matrices, we have proposed a second procedure based on an approximation. An implementation of both methods is available in the \texttt{splitFlip} package \citep{splitFlip} in \texttt{R}.

Our method is extremely flexible, allowing different selection procedures and several combining functions. Particular attention is to be paid to the choice of the selection procedure, as it must fulfill the method's assumptions. We suggest using the Lasso \citep{lassotib} with a suitable calibration of the $\lambda$ parameter, so that enough variables are selected for the screening property to be likely. More research is needed on the properties of combining functions for the individual statistics; we expect that different functions will have different power properties, and will perform best in different scenarios. As the methods are asymptotic, their behavior should be further explored with finite sample size to analyze in which cases asymptotic properties still hold. Moreover, the test of \citet{score} and \citet{finosflip} that our method builds on is robust against some model misspecifications; hence it would be of interest to assess if the method maintains such robustness.


\bibliography{biblio}


\newpage
\appendix

\section{Algorithmic implementation}\label{appen:algs}
We provide an outline and pseudocode for the relevant procedures presented in this manuscript: the Multisplit method \citep{multisplit}, as well as the proposed exact and approximate methods.

\subsection{Multisplit}\label{alg_multi}
Algorithm \ref{algorithm:meinshausen} implements the Multisplit method \citep{multisplit} introduced in Section \ref{descrmultisplit}, which provides adjusted p-values $p_1,\ldots,p_m$ for all variables.

\begin{algorithm}
 \caption{\label{algorithm:meinshausen} Multisplit method to compute $p_j$ for each $j\in\Mset$.}
\SetAlgoLined

\KwData{$\yvec\in\Rset^n$; $\xmat\in\Rset^{n\times m}$; Q}
\KwResult{$p_1,\ldots,p_m$}

$\mathbf{W} = q\times m$ null matrix\;

\For{$q=1,\ldots,Q$}{
   randomly split $\{1,\ldots,n\}$ into $\Din$ and $\Dout$\;
   use $\yvec_{\Din}$ and $\xmat_{\Din,\Mset}$ to select variables $\Aset^q\subseteq\Mset$\;
   \For{$j=1,\ldots,m$}{
      \uIf{$j\in \Aset^q$}{
         $\tilde{p}_j^q =$ raw p-value computed via OLS estimation with $\yvec_{\Dout}$ and $\xmat_{\Dout,\Aset^q}$\;
      }\Else{
         $\tilde{p}_j^q = 1$\;
      }
      $W_{qj}=p_j^q$ computed as in \eqref{def:adjunaggregated}\;
   }
}

\For{$j=1,\ldots,m$}{
   compute $p_j$ as in \eqref{def:adjaggregated} using the $j$-th column of $\mathbf{W}$\;
}

\Return $p_1,\ldots,p_m$\;
\end{algorithm}


\subsection{Exact method}\label{alg_splitflipe}
Algorithm \ref{algorithm:exact} implements the exact method of Section \ref{rbmulti} that uses $B$ random sign-flipping transformations and $Q$ splits to compute the test statistics $\Us^1,\ldots,\Us^B$ for all variables $j\in\Mset$. Results are returned as a $B\times m$ matrix of test statistics, where columns correspond to variables and rows to transformations.

\begin{algorithm}
 \caption{\label{algorithm:exact} Method to compute $\Us^b$ for $j\in\Mset$ and $b\in\{1,\ldots,B\}$.}
\SetAlgoLined

\KwData{$\yvec\in\Rset^n$; $\xmat\in\Rset^{n\times m}$; B; Q}
\KwResult{$\mathbf{G}$ ($B\times m$ matrix with $G_{bj}=\Us^b$)}

$\mathbf{G}=B\times m$ null matrix\;
$\fmat[1]=n\times n$ identity matrix\;
$\fmat[2],\ldots,\fmat[B]=n\times n$ diagonal matrices with elements independently and uniformly drawn from $\{-1,1\}$\;
Queue = empty list\;

\For{$q=1,\ldots,Q$}{
   randomly split $\{1,\ldots,n\}$ into $\Din$ and $\Dout$\;
   use $\yvec_{\Din}$ and $\xmat_{\Din,M}$ to select variables $\Aset^q\subseteq\Mset$\;
   add $(\Dout,\Aset^q)$ to Queue\;
}

\For{$j=1,\ldots,m$}{
   Rs = empty list\;
   
   \For{$q=1,\ldots,Q$}{
      $(\Dout,\Aset^q)=q$-th elements of Queue\;
      \If{$j\in \Aset^q$}{
         compute $\re^q$ as in \eqref{def:splitresmat}\;
         add $\re^q$ to Rs\;}
   }
   
   \lIf{Rs is empty}{$\mathbf{G}_{\{1,\ldots,B\},j}=(0,\ldots,0)$}
   
   \For{$b=1,\ldots,B$}{
      compute $\Ue^b$ as in \eqref{def:Ue} using elements of Rs\;
      $G_{bj}=$ standardization of $\Ue^b$ as in \eqref{def:Ts}\;
   }
}

\Return $\mathbf{G}$\;
\end{algorithm}

The following lemma shows the worst-case computational complexity and memory usage of the algorithm. The complexity is polynomial in the sample size $n$, and linear both in the number $Q$ of splits and in the number $B$ of transformations. The memory usage is quadratic in $n$ and linear in $Q$.


\begin{lemma}\label{L:complexity_exact}
In the worst case, Algorithm \ref{algorithm:exact} (excluding the variable selection procedure) has computational complexity of order $n^4QB$, and memory usage of order $n^2Q$.
\end{lemma}



\subsection{Approximate method}\label{alg_splitflipa}
Algorithm \ref{algorithm:approximate} implements the method of Section \ref{rbmultia}, and represents an approximation for the procedure of Algorithm \ref{algorithm:exact}. It relies on $B$ random sign-flipping transformations and $Q$ splits to compute the test statistics $\Vs^1,\ldots,\Vs^B$ for all variables $j\in\Mset$. Results are returned as a $B\times m$ matrix of test statistics, where columns correspond to variables and rows to transformations.

\begin{algorithm}
 \caption{\label{algorithm:approximate} Method to compute $\Vs^b$ for $j\in\Mset$ and $b\in\{1,\ldots,B\}$.}
\SetAlgoLined

\KwData{$\yvec\in\Rset^n$; $\xmat\in\Rset^{n\times m}$; B; Q}
\KwResult{$\mathbf{G}$ ($B\times m$ matrix with $G_{bj}=\Vs^b$)}

$\mathbf{G}=B\times m$ null matrix\;
$\fmat[1]=n\times n$ identity matrix\;
$\fmat[2],\ldots,\fmat[B]=n\times n$ diagonal matrices with elements independently and uniformly drawn from $\{-1,1\}$\;
Queue = empty list\;

\For{$q=1,\ldots,Q$}{
   randomly split $\{1,\ldots,n\}$ into $\Din$ and $\Dout$\;
   use $\yvec_{\Din}$ and $\xmat_{\Din,M}$ to select variables $\Aset^q\subseteq\Mset$\;
   add $(\Dout,\Aset^q)$ to Queue\;
}

\For{$j=1,\ldots,m$}{
   $\reb=n\times n$ null matrix\;
   
   \For{$q=1,\ldots,Q$}{
      $(\Dout,\Aset^q)=q$-th elements of Queue\;
      \If{$j\in \Aset^q$}{
         compute $\re^q$ as in \eqref{def:splitresmat}\;
         $\reb=\reb+\re^q$\;}
   }
   
   \lIf{$\reb$ is null}{$\mathbf{G}_{\{1,\ldots,B\},j}=(0,\ldots,0)$}
   
   \For{$b=1,\ldots,B$}{
      compute $\Ve^b$ as in \eqref{def:Ve} using $\reb$\;
      $G_{bj}=$ standardization of $\Ve^b$ as in \eqref{def:Ts}\;
   }
}
\Return $\mathbf{G}$\;
\end{algorithm}

As shown in the following lemma, the computational complexity and the memory usage are lower than those of Algorithm \ref{algorithm:exact}.

\begin{lemma}\label{L:complexity_approx}
In the worst case, Algorithm \ref{algorithm:approximate} (excluding the variable selection procedure) has computational complexity of order $n^4Q + n^3B$, and memory usage of order $n^2$.
\end{lemma}


\section{Simulations}\label{appen:simmulti}
In this section, we give additional information on the simulations of Section \ref{multisims}. We provide results for the simulated matrix, as well as the computation time for Section \ref{multisims2}.

\subsection{Approximate and exact}\label{appen:simmulti1}
\begin{figure}
\centering
\includegraphics[width=\textwidth]{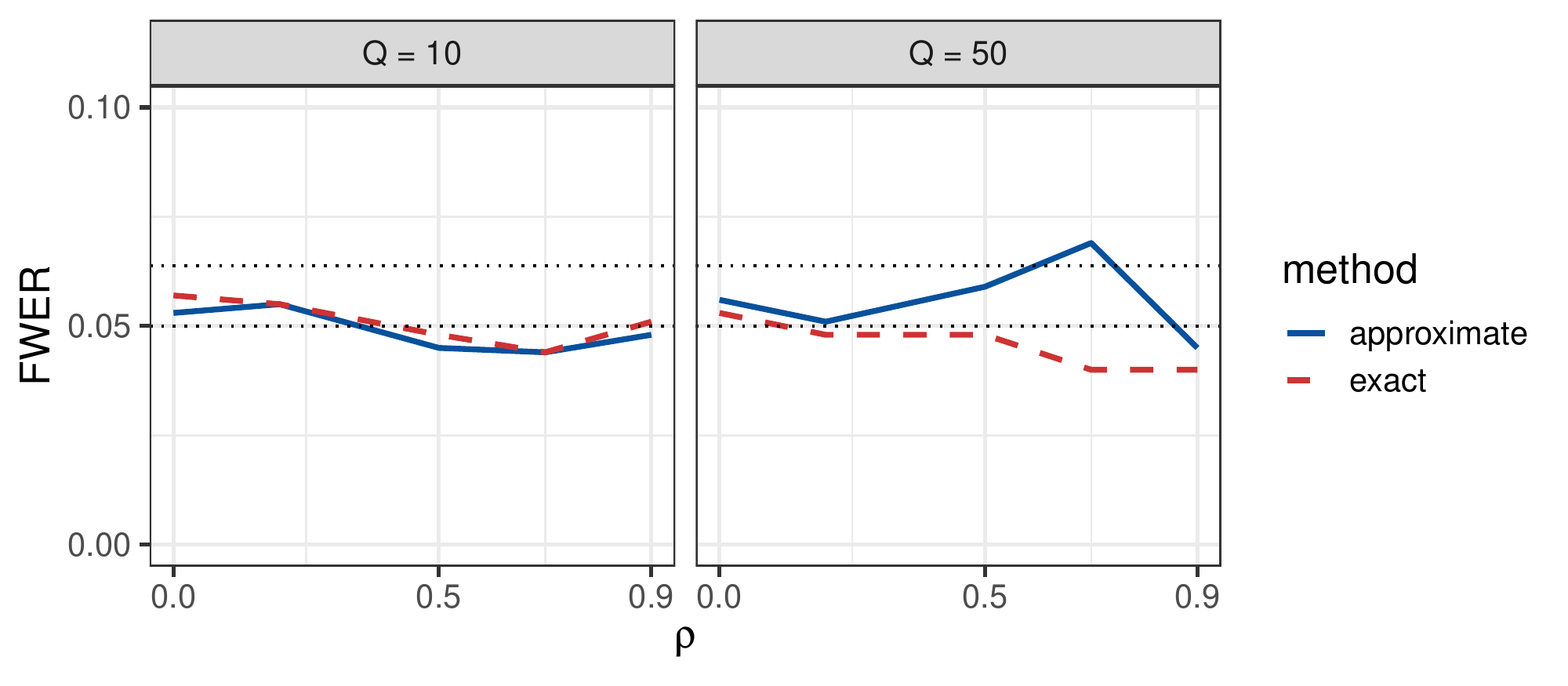}
\caption{Simulated design matrix with $m=100$: FWER by covariance parameter $\rho$, for the approximate and exact methods using $Q$ splits. The dotted lines correspond to the significance level $\alpha=0.05$ and an upper bound ($\alpha$ plus two standard deviations, approximately 0.063).}
\label{Plot:EAfwer}
\end{figure}

\begin{figure}
\centering
\includegraphics[width=\textwidth]{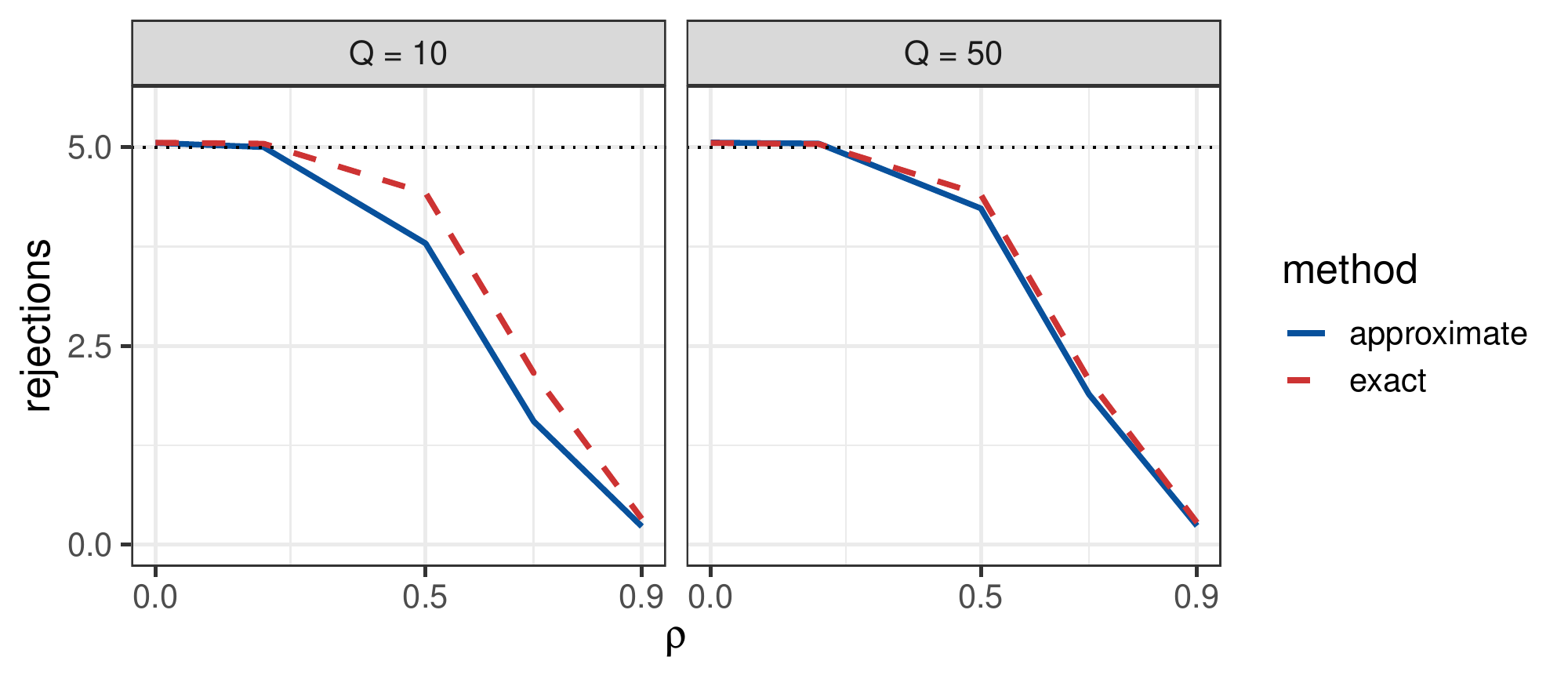}
\caption{Simulated design matrix with $m=100$: number of rejections by covariance parameter $\rho$, for the approximate and exact methods using $Q$ splits. The dotted line denotes the true number of active variables.}
\label{Plot:EAtdp}
\end{figure}

Figures \ref{Plot:EAfwer} and \ref{Plot:EAtdp} show the FWER and the total number of rejections obtained using the simulated design matrix. The exact method always controls the FWER, as it never exceeds the significance level $\alpha$ by more than two standard deviations; the FWER given by the approximate method exceeds this threshold only in the setting with $\rho=0.7$ and $Q=50$. In terms of number of rejections, the approximate method is close to the exact, especially when the number of splits is high.

\subsection{Approximate and Multisplit}\label{appen:simmulti2}
\begin{figure}
\centering
\includegraphics[width=\textwidth]{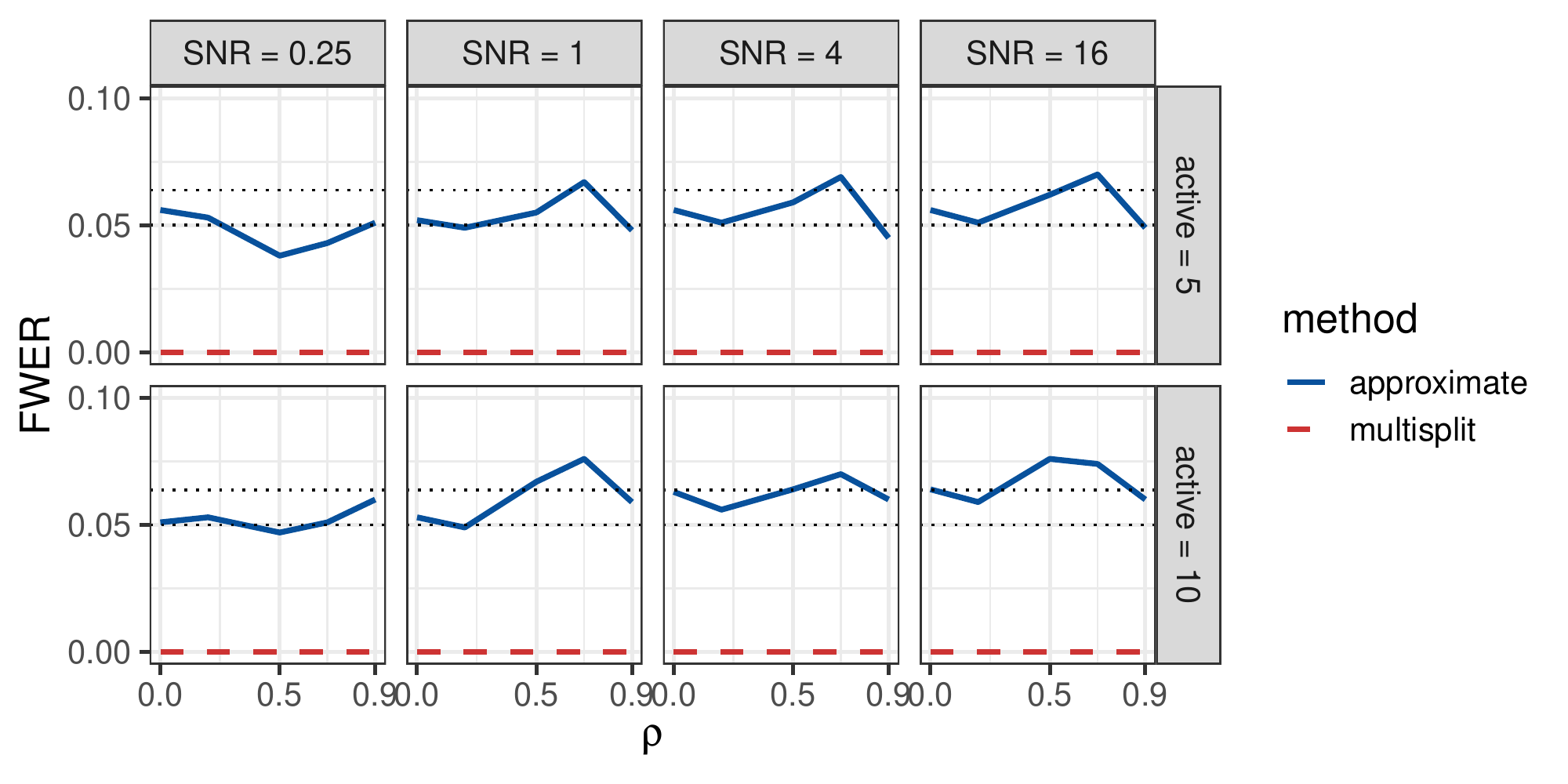}
\caption{Simulated design matrix with $m=100$: FWER by covariance parameter $\rho$, for the approximate method and the Multisplit. \textit{Active} and SNR denote the true number of active variables and the signal-to-noise ratio. The dotted lines correspond to the significance level $\alpha=0.05$ and an upper bound ($\alpha$ plus two standard deviations, approximately 0.063).}
\label{Plot:MAfwer1}
\end{figure}

\begin{figure}
\centering
\includegraphics[width=\textwidth]{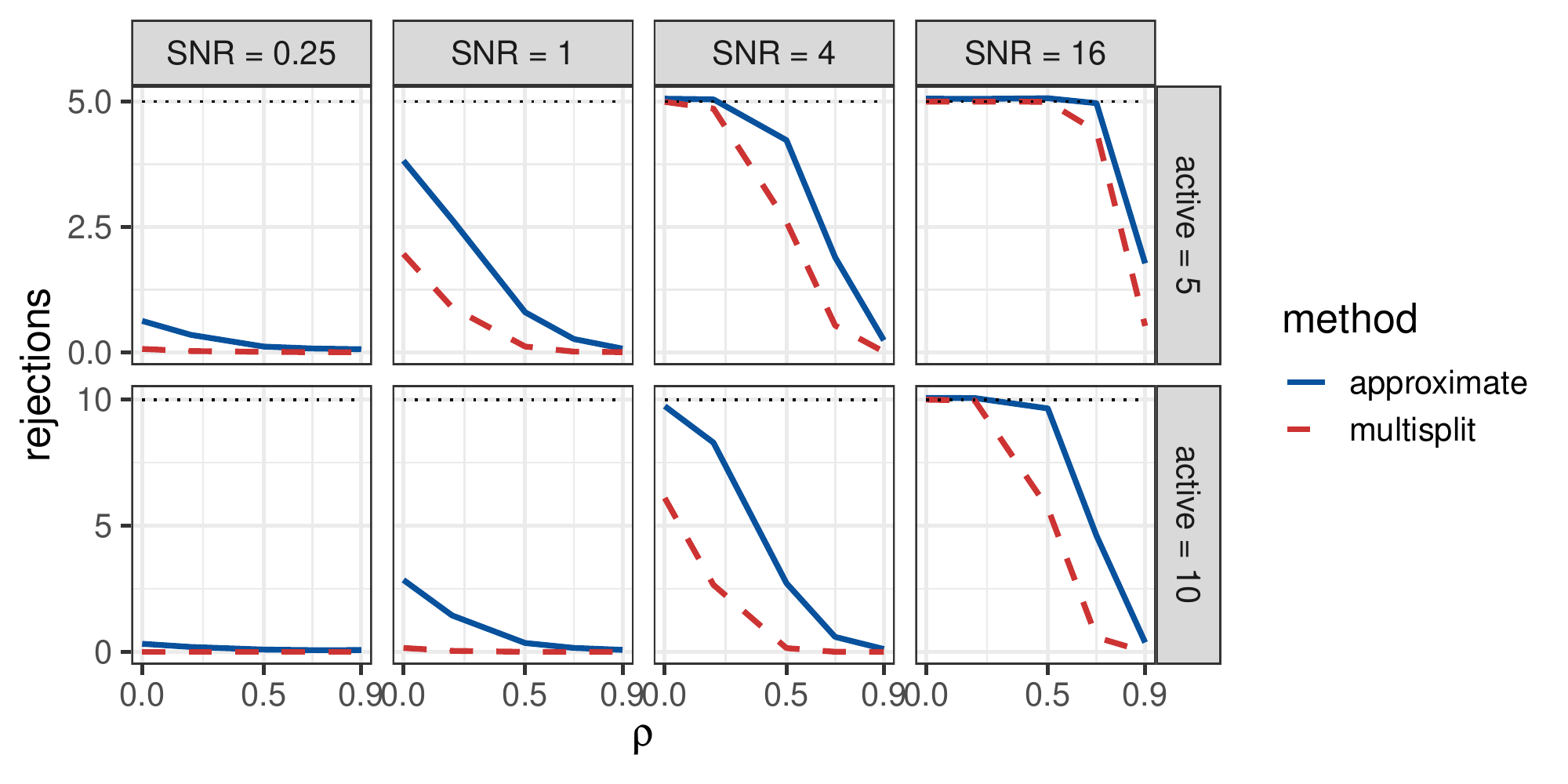}
\caption{Simulated design matrix with $m=100$: number of rejections by covariance parameter $\rho$, for the approximate method and the Multisplit. \textit{Active} and SNR denote the number of active variables and the signal-to-noise ratio. The dotted line corresponds to \textit{active}.}
\label{Plot:MAtdp1}
\end{figure}

\begin{figure}
\centering
\includegraphics[width=\textwidth]{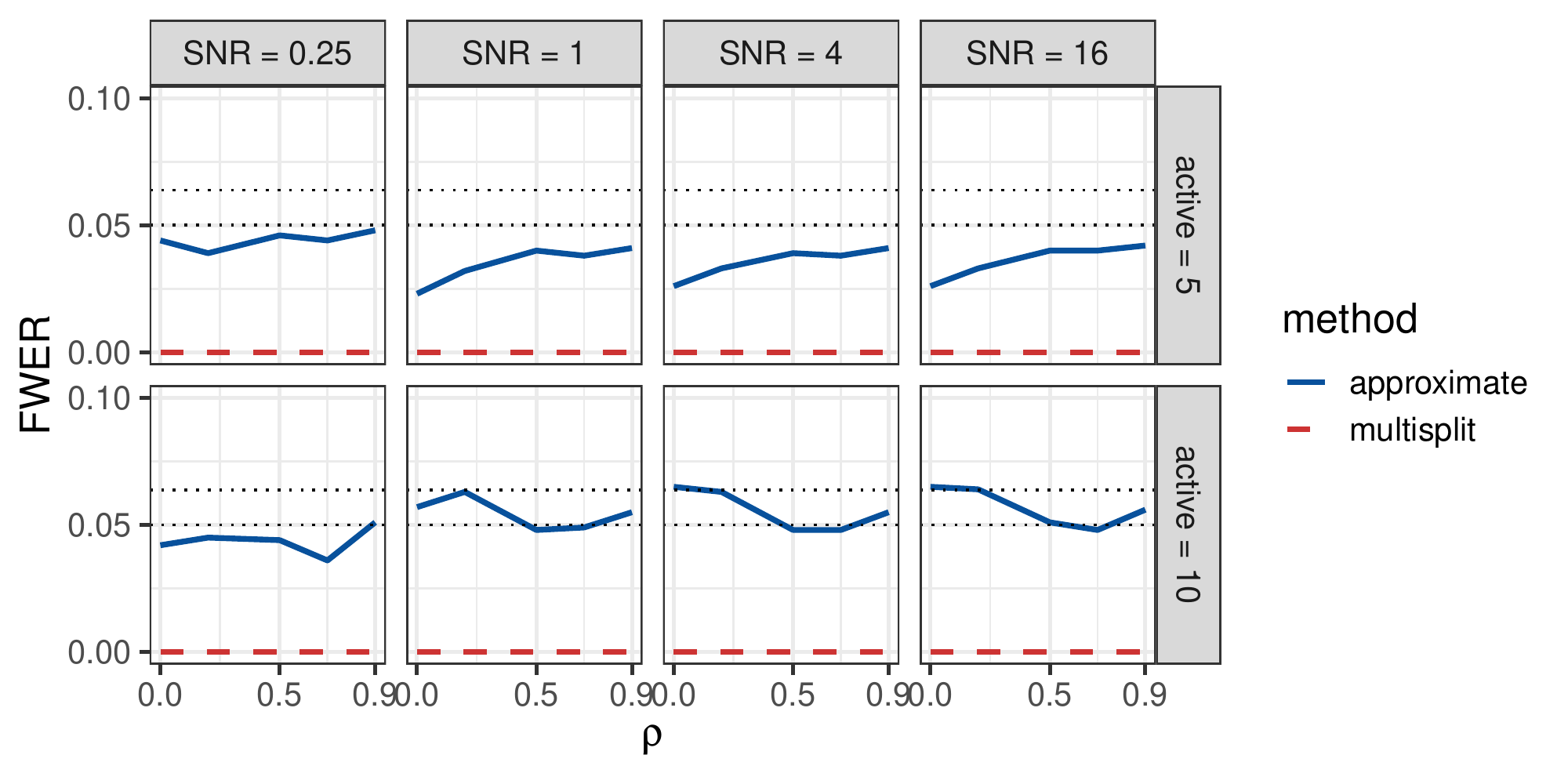}
\caption{Simulated design matrix with $m=1000$: FWER by covariance parameter $\rho$, for the approximate method and the Multisplit. \textit{Active} and SNR denote the true number of active variables and the signal-to-noise ratio. The dotted lines correspond to the significance level $\alpha=0.05$ and an upper bound ($\alpha$ plus two standard deviations, approximately 0.063).}
\label{Plot:MAfwer2}
\end{figure}

\begin{figure}
\centering
\includegraphics[width=\textwidth]{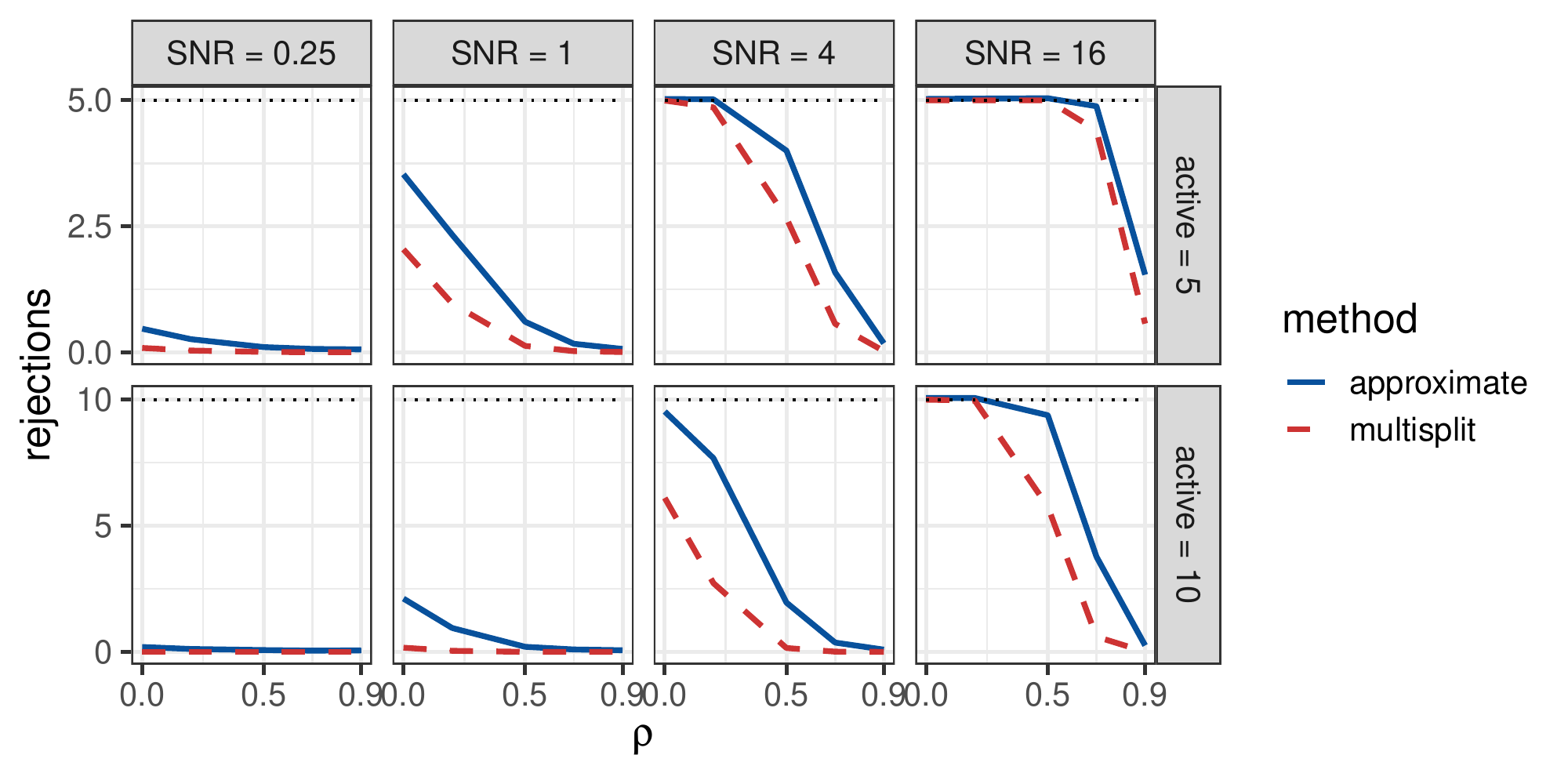}
\caption{Simulated design matrix with $m=1000$: number of rejections by covariance parameter $\rho$, for the approximate method and the Multisplit. \textit{Active} and SNR denote the number of active variables and the signal-to-noise ratio. The dotted line corresponds to \textit{active}.}
\label{Plot:MAtdp2}
\end{figure}

Figures \ref{Plot:MAfwer1} and \ref{Plot:MAtdp1} show results for the simulated design matrix with $m=100$. The proposed method appears to control the FWER when the covariance parameter $\rho$ is not too high, or the signal-to-noise ratio SNR is particularly low. Among the scenarios where the FWER is controlled, the method is always more powerful than the Multisplit, with greatest differences when $\rho$ and SNR are low. Results for $m=1000$ (Figures \ref{Plot:MAfwer2} and \ref{Plot:MAtdp2}) display an analogous behavior.

Finally, Table \ref{Table:MA_time} contains the maximum computation time over the different settings for both the simulated and real design matrices.

\begin{table}
\centering
\caption{Simulated and real design matrices with $m$ variables: maximum computation time (s) for the approximate method and the Multisplit.}
\label{Table:MA_time}
\begin{tabular}{cccc}
\toprule
  & \multicolumn{2}{c}{simulated} & real\\
$m$  & 100 & $1{,}000$ & $4{,}088$\\
\midrule
approximate  & 34.2 & 137.7 & 58.8\\
Multisplit  & 0.08 & 0.33 & 1.3\\
\bottomrule
\end{tabular}
\end{table}


\subsection{Oracle and Lasso}\label{appen:simmulti3}
\begin{figure}
\centering
\includegraphics[width=\textwidth]{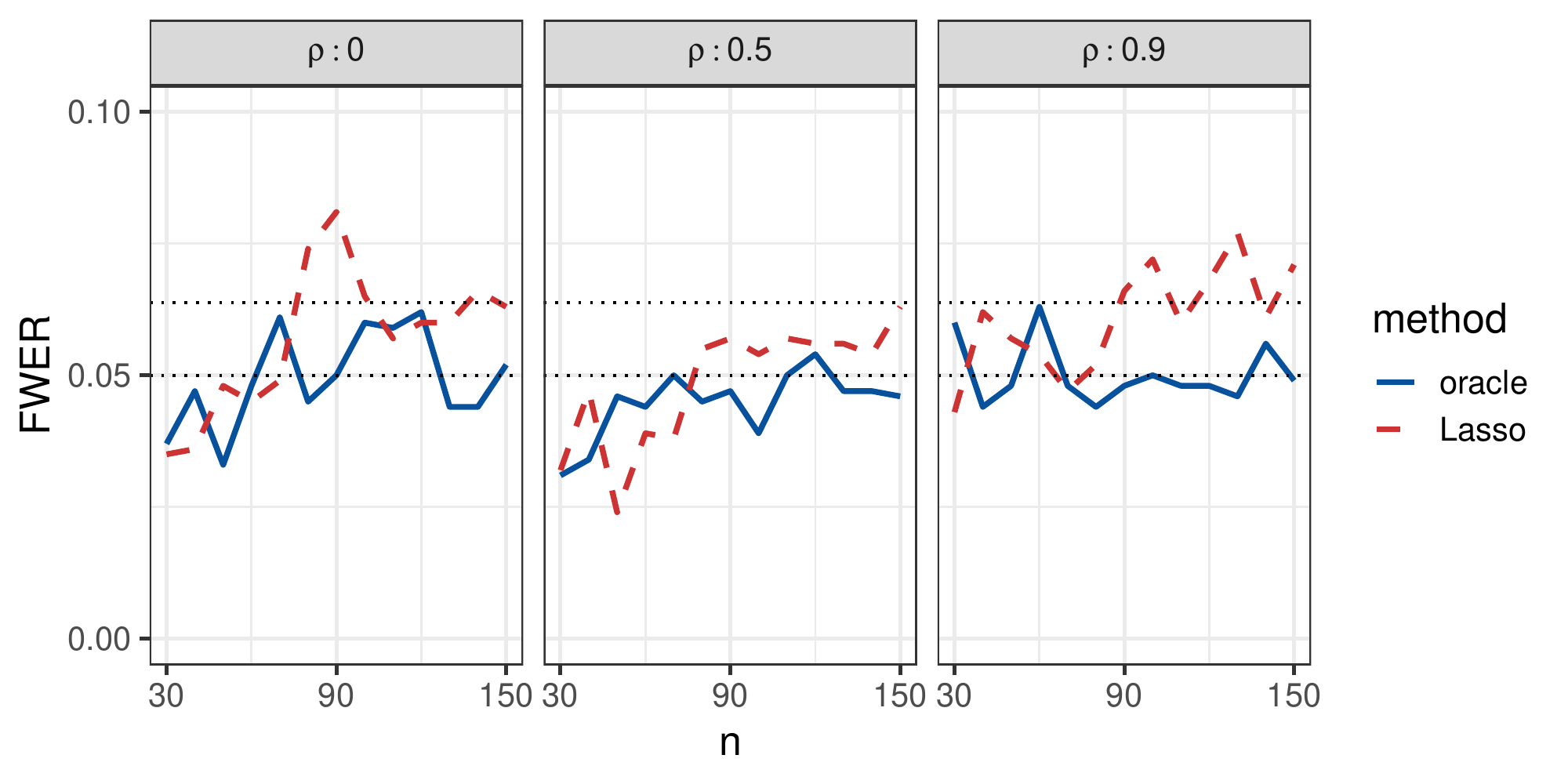}
\caption{Simulated design matrix with $m=100$: FWER by sample size $n$, for the approximate method using oracle selection and Lasso. $\rho$ denotes the covariance parameter. The dotted lines correspond to the significance level $\alpha=0.05$ and an upper bound ($\alpha$ plus two standard deviations, approximately 0.063).}
\label{Plot:LOfwer}
\end{figure}

\begin{figure}
\centering
\includegraphics[width=\textwidth]{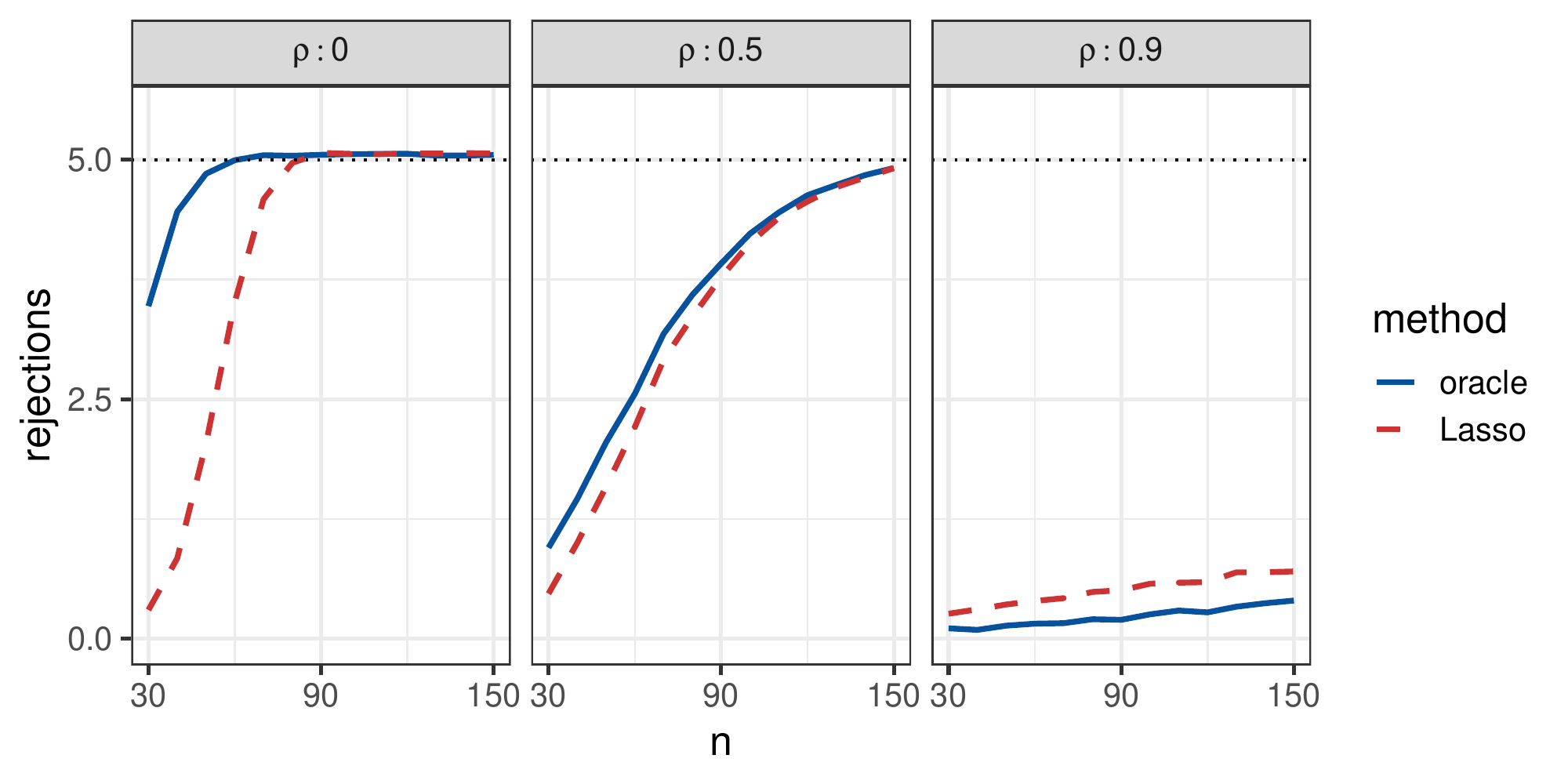}
\caption{Simulated design matrix with $m=100$: number of rejections by sample size $n$, for the approximate method using oracle selection and Lasso. $\rho$ denotes the covariance parameter. The dotted line corresponds to the true number of active variables.}
\label{Plot:LOtdp}
\end{figure}

Figure \ref{Plot:LOfwer} and Figure \ref{Plot:LOtdp} show results for the comparison between oracle selection and Lasso using the simulated design matrix with $n\in\{30,40,\ldots,150\}$. In the considered settings, the FWER is always controlled by the oracle but not by the Lasso, which fails in some scenarios both with low and high covariance parameter $\rho$ . Moreover, in the cases where both methods control the FWER, as expected the oracle is always at least as powerful as the Lasso. This difference is more noticeable when $n$ and $\rho$ are low.

\section{Proofs}\label{multiproofs}


\subsection{Projection matrices.}\label{projprop}
Here we recall some properties of projection matrices, which will be used within proofs. For any $j\in\Mset$, define the hat matrix
\[\he = \ze (\ze^\top\ze)^{-1}\ze^\top\in\Rset^{n\times n} \]
and the residual maker matrix
\[\re = \mathbf{I}-\he\in\Rset^{n\times n}\]
as given in \eqref{def:rmmatrix}. The following properties hold.
\begin{enumerate}
\item If a matrix $\mathbf{R}\in\mathbb{R}^{n\times n}$ is symmetric ($\mathbf{R}^\top = \mathbf{R}$) and idempotent ($\mathbf{R}\mathbf{R}=\mathbf{R}$), then it is a projection matrix. As a consequence, it is positive semi-definite, i.e., $\mathbf{z}^\top \mathbf{R} \mathbf{z}\geq 0$ for any $\mathbf{z}\in\mathbb{R}^n$.
\item $\he$ and $\re$ are projection matrices with $\he\ze =\ze$ and $\re\ze =\mathbf{0}$.
\item All diagonal elements of $\he=(h_{k\ell})$ converge to zero, i.e.,
\[\lim_{\mylim}\max_k h_{kk}=0 \]
(see \citet{huber}, Proposition 7.1).
\end{enumerate}


\paragraph{Proposition \ref{P:signflip}.}
\textit{The test that rejects $H_j$ when $|\Te^1| > |\Te^{(\lceil (1-\alpha)B\rceil)}|$ is asymptotically an $\alpha$-level test for any $j\in\Mset$. For finite $n$, it may be anti-conservative as
\[\myvar(\Te^1)\geq\myvar(\Te^b)\qquad (b\in\{1,\ldots,B\}).\]}

\begin{proof}
Proof of the first part of the proposition in a more general case is in \citet{score} (see Theorem 2). We briefly recall the main steps in our notation. Fix any $j\in\Mset$, and consider any couple of transformations $b,c\in\{1,\ldots,B\}$. Assume that $H_j$ is true, so that
\[\yvec=\ze\bevec_{-j}+\epsvec,\qquad \epsvec\sim\multinormal (\mathbf{0},\sigma^2 \mathbf{I})\]
and, by the properties of $\re$ given in Section \ref{projprop},
\begin{align}
\Te^b=\tjo^\top \yvec = \frac{1}{\sqrt{n}}\ve^\top\re \fmat\re (\ze\bevec_{-j}+\epsvec) = \tjo^\top\epsvec . \label{vyeqve}
\end{align}
First the Authors prove that
\begin{align}
\Te^b=\Te^{*b}+o_P(1),\qquad \Te^{*b}=\frac{1}{\sqrt{n}}\ve^\top\re \fmat\epsvec=\tjo[1]^\top \fmat\epsvec \label{testar0}
\end{align}
as $\mylim$, and so the $B$-dimensional vectors
\[\Tevect=(\Te^1,\ldots,\Te^B)^\top,\qquad \Tevect^*=(\Te^{*1},\ldots,\Te^{*B})^\top\]
are asymptotically equivalent. Indeed, since the diagonal elements of $\fmat[2],\ldots,\fmat[B]$ are independently and uniformly drawn from $\{-1,1\}$,
\begin{align*}
\Te^{*b} - \Te^b &= \frac{1}{\sqrt{n}}\ve^\top\re \fmat\he\epsvec=\tjo[1]^\top \fmat\he\epsvec\\
\myexp(\Te^{*b} - \Te^b)&=0\\
\myvar (\Te^{*b} - \Te^b)&=
\sigma^2\tjo[1]^\top\myexp(\fmat\he \fmat)\tjo[1]=
\sigma^2\tjo[1]^\top\text{diag}(\he)\tjo[1]\leq\sigma^2\|\tjo[1]\|^2 \max_k h_{kk}\longrightarrow 0
\end{align*}
(see Section \ref{projprop}). Now focus on $\Tevect^*$, and observe that
\begin{align*}
\myexp(\Te^{*b})&=0\\
\mycov (\Te^{*b},\Te^{*c})&=\frac{\sigma^2}{n}\tjo[1]^\top\myexp (\fmat\fmat[c])\tjo[1]=
\begin{cases}
\sigma^2 \|\tjo[1]\|^2\quad\text{if}\quad b=c\\
0\quad\text{otherwise.}
\end{cases}
\end{align*}
Therefore
\[\Tevect^*\sim\multinormal[B]\left(\mathbf{0}, \sigma^2 \|\tjo[1]\|^2 \mathbf{I}\right) \]
and so
\[\Tevect,\Tevect^*\limdistr \mathbf{Z}\sim\multinormal[B]\left(\mathbf{0},\xi^2 \mathbf{I}\right),\qquad \xi^2=\sigma^2\lim_{\mylim}\|\tjo[1]\|^2.\]
As a consequence, the statistics $\Te^1,\ldots,\Te^B$ converge to i.i.d.~random variables, and so do the statistics $|\Te^1|,\ldots,|\Te^B|$. As high values of $|\Te^1|$ correspond to evidence against $H_j$,
\[\lim_{\mylim}P\left(|\Te^1| > |\Te^{(\lceil (1-\alpha)B\rceil)}|\right) = \frac{\lfloor\alpha B\rfloor}{B}\leq\alpha\]
(Lemma 1 in \citet{score}).

Proof of the second part of the Proposition is in \citet{finosflip}. To show that $\myvar (\Te) \geq \myvar (\Te^b)$, it is sufficient to observe that
\begin{align*}
\myvar (\Te^b) &= \frac{\sigma^2}{n}\ve^\top\re\myexp(\fmat\re \fmat)\re\ve= 
\sigma^2\tjo[1]^\top \myexp(\fmat\re \fmat)\tjo[1]\\
\myvar (\Te^1) - \myvar (\Te^b) &= \sigma^2 \myexp\left(\tjo[1]^\top \fmat\he \fmat\tjo[1]\right) \geq 0
\end{align*}
since $\fmat\he \fmat$ is a projection matrix, and so positive semi-definite (see Section \ref{projprop}).
\end{proof}

\paragraph{Theorem \ref{T:signflip2}.}
\textit{The test that rejects $H_j$ when $|\Ts^1| > |\Ts^{(\lceil (1-\alpha)B\rceil)}|$ is an $\alpha$-level test for any $j\in\Mset$.}

\begin{proof}
Proof of the theorem for the more general case of generalized linear models is in \citet{finosflip}, and we recall it here for the case of the linear model. Fix any $j\in\Mset$, and consider any couple of transformations $b,c\in\{1,\ldots,B\}$. Assume that $H_j$ is true, so that from \eqref{vyeqve} we have
\[\Ts^b = \tj^\top Y =\tj^\top\varepsilon .\]
Then consider the $B$-dimensional vector $\Tsvect=(\Ts^1,\ldots,\Ts^B)^\top$. Similarly to the proof of Proposition \ref{P:signflip},
\begin{align*}
\myexp(\Ts^b)&=0\\
\mycov (\Ts^b,\Ts^c)&=\sigma^2\myexp (\tj^\top \tj[c]) =
\begin{cases}
\sigma^2\quad\text{if}\quad b=c\\
0\quad\text{otherwise.}
\end{cases}
\end{align*}
Hence $\Tsvect\sim\mathcal{N}_B(0,\sigma^2 I)$ is a vector of i.i.d.~random variables, and so also $|\Ts^1|,\ldots,|\Ts^B|$ are i.i.d.~random variables. As high values of $|\Te^1|$ correspond to evidence against $H_j$, then
\[P\left(|\Ts^1| > |\Ts^{(\lceil (1-\alpha)B\rceil)}|\right) = \frac{\lfloor\alpha B\rfloor}{B}\leq\alpha\]
by the Monte Carlo testing principle \citep{lehmann}.
\end{proof}


\paragraph{Lemma \ref{L:combinevars}.}
\textit{The test that rejects $H_\Sset$ when $\Ts[\Sset]^1 > \Ts[\Sset]^{(\lceil (1-\alpha)B\rceil)}$ is asymptotically an $\alpha$-level test for any non-empty $\Sset\subseteq\Mset$.}


\begin{proof}
Fix any non-empty set $\Sset =\{j_1,\ldots,j_s\}\subseteq\Mset$, and assume that $H_\Sset$ is true. As each $H_j$ with $j\in\Sset$ is individually true, we can write
\[\yvec=\ze[j_1]\bevec_{-j_1}+\epsvec=\ldots =\ze[j_s]\bevec_{-j_s}+\epsvec,\qquad \epsvec\sim\multinormal (\mathbf{0},\sigma^2 \mathbf{I}).\]
Consider any couple of variables $j,h\in \Sset$ and any couple of transformations $b,c\in\{1,\ldots,B\}$. Similarly to the proof of Proposition 1, define the $sB$-dimensional vector
\[\Tevect[\Sset]=(\Te[j_1]^1,\ldots,\Te[j_1]^B,\ldots,\Te[j_s]^1,\ldots,\Te[j_s]^B)^\top,\qquad \Te^b=\tjo^\top\yvec=\tjo^\top\epsvec.\]
From \eqref{testar0}, $\Tevect[\Sset]$ is asymptotically equivalent to 
\[\Tevect[\Sset]^* = (\Te[j_1]^{*1},\ldots,\Te[j_1]^{*B},\ldots,\Te[j_s]^{*1},\ldots,\Te[j_s]^{*B})^\top , \qquad  \Te^{*b}=\tjo[1]^\top \fmat\epsvec \]
as $\mylim$. Moreover, $\Tevect[\Sset]^*$ follows a multivariate normal distribution with
\begin{align*}
\myexp(\Te^{*b})&=0\\
\mycov (\Te^{*b},\Te[h]^{*c})&=\sigma^2\tjo[1]^\top\myexp (\fmat \fmat[c]) \mathbf{t}_{h1}=
\begin{cases}
\sigma^2 \tjo[1]^\top \mathbf{t}_{h1}\quad\text{if}\quad b=c\\
0\quad\text{otherwise.}
\end{cases}
\end{align*}
Then
\[\Tevect[\Sset],\Tevect[\Sset]^*\limdistr \mathbf{Z}\sim\multinormal[sB]\left(\mathbf{0},\boldsymbol{\Xi}\otimes \mathbf{I}\right)\]
where $\otimes$ denotes the Kronecker product and
\begin{align*}
\mathbf{I}\in\Rset^{B\times B},\qquad \boldsymbol{\Xi}=(\xi_{k\ell})\in\Rset^{s\times s},\qquad 
\xi_{k\ell}=\sigma^2 \lim_{\mylim}\mathbf{t}_{j_k 1}^\top \mathbf{t}_{j_\ell 1}.
\end{align*}
Equivalently, we can say that
\begin{align*}
\begin{pmatrix}
T_{j_1}^1 & \ldots & T_{j_s}^1\\
\vdots &  & \vdots\\
T_{j_1}^B & \ldots & T_{j_s}^B
\end{pmatrix}
\limdistr Z^{\prime}\sim\matrixnormal[s\times B]\left(\mathbf{0},\mathbf{I},\boldsymbol{\Xi}\right)
\end{align*}
where $\matrixnormal[s\times B]$ denotes the matrix normal distribution.

Hence the $B$ vectors of effective scores $(\Te[j_1]^1,\ldots,\Te[j_s]^1),\ldots, (\Te[j_1]^B,\ldots,\Te[j_s]^B)$ converge to i.i.d.~random vectors, and so do the $B$ vectors of the absolute values of standardized scores $(|\Ts[j_1]^1|,\ldots,|\Ts[j_s]^1|),\ldots,$ $(|\Ts[j_1]^B|,\ldots,|\Ts[j_s]^B|)$. Therefore the combinations of their elements $\Ts[\Sset]^1,\ldots,\Ts[\Sset]^B$ defined in \eqref{def:combTs} converge to i.i.d.~random variables. Moreover, for each variable $j$ high values of $|\Ts^1|$ correspond to evidence against $H_j$, and $g$ is increasing in each argument. Therefore high values of $\Ts[\Sset]^1$ correspond to evidence against $H_\Sset$. From \citet{score} (see Lemma 1),
\[\lim_{\mylim}P\left(\Ts[\Sset]^1 > \Ts[\Sset]^{(\lceil (1-\alpha)B\rceil)}\right) = \frac{\lfloor\alpha B\rfloor}{B}\leq\alpha.\]
\end{proof}

\paragraph{Theorem \ref{T:multi_exact}.}
\textit{The test that rejects $H_\Sset$ when $\Us[\Sset]^1 > \Us[\Sset]^{(\lceil (1-\alpha)B\rceil)}$ is asymptotically an $\alpha$-level test for any non-empty $\Sset\subseteq\Mset$.}

\begin{proof}
Fix any non-empty set $\Sset =\{j_1,\ldots,j_s\}\subseteq\Mset$, and assume that $H_\Sset$ is true. Then consider any split $q\in\{1,\ldots,Q\}$, any couple of variables $j,h\in\Sset$ and any couple of transformations $b,c\in\{1,\ldots,B\}$. Moreover, suppose that the variable selection procedure selects all active variables; by Assumption \ref{A:screening}, this is true at least asymptotically, so this assumption does not affect asymptotic results. As $H_j$ is true and all active variables are contained in $\Aset^q$, we can write
\begin{align*}
\yvec_{\Dout} = \xmat_{-j;\Dout,\Aset^q} \bevec_{-j; \Aset^q} + \epsvec_{\Dout},\qquad\epsvec_{\Dout}\sim\multinormal[n/2](\mathbf{0},\sigma^2 \mathbf{I})
\end{align*}
considering only observations in $\Dout$ and variables in $\Aset^q$. The effective score \eqref{def:Te} for this model is
\begin{align*}
\Te^{qb}=
\begin{cases}
\frac{1}{\sqrt{n}}\xmat_{j;\Dout}^\top \mathbf{R}_{-j;\Dout,\Dout}^q \mathbf{F}_{b;\Dout,\Dout} \mathbf{R}_{-j;\Dout,\Dout}^q \yvec_{\Dout}\quad\text{if}\quad j\in \Aset^q\\
0\quad\text{otherwise}
\end{cases}
= \frac{1}{\sqrt{n}}\ve^\top \re^q \fmat \re^q\yvec .
\end{align*}
Indeed, the matrix $\re^q$ given in \eqref{def:rmmatrix} has non-null elements only corresponding to observations in $\Dout$ if $j$ is selected, and is null otherwise. From \eqref{testar0}, as $\mylim$ we have
\begin{align*}
\Te^{qb}=\Te^{*qb}+o_P(1),\qquad \Te^{*qb}=\frac{1}{\sqrt{n}}\ve^\top\re^q \fmat\varepsilon
\end{align*}
and, summing over the splits,
\begin{align}
\Ue^b=\sum_{q=1}^Q\Te^{qb}=\ujo^\top\epsvec=\Ue^{*b}+o_P(1),\qquad \Ue^{*b}=\sum_{q=1}^Q\Te^{*qb}=\ujo[1]^\top \fmat\epsvec .  \label{def:testarq}
\end{align}
Therefore the two $sB$-dimensional vectors
\begin{align*}
\Uevect[\Sset]&=(\Ue[j_1]^1,\ldots,\Ue[j_1]^B,\ldots,\Ue[j_s]^1,\ldots,\Ue[j_s]^B)^\top\\
\Uevect[\Sset]^*&=(\Ue[j_1]^{*1},\ldots,\Ue[j_1]^{*B},\ldots,\Ue[j_s]^{*1},\ldots,\Ue[j_s]^{*B})^\top
\end{align*}
are asymptotically equivalent.

The second part of the proof is analogous the the proof of Lemma \ref{L:combinevars}. Indeed, it is sufficient to observe that $\Uevect[\Sset]^*$ follows a multivariate normal distribution with
\begin{align*}
\myexp(\Ue^{*b})&=0\\
\mycov (\Ue^{*b},\Ue[h]^{*c})&=\sigma^2\ujo[1]^\top\myexp (\fmat\fmat[c]) \mathbf{u}_{h1}=
\begin{cases}
\sigma^2 \ujo[1]^\top \mathbf{u}_{h1}\quad\text{if}\quad b=c\\
0\quad\text{otherwise.}
\end{cases}
\end{align*}
Therefore
\[\Uevect[\Sset],\Uevect[\Sset]^*\limdistr \mathbf{Z}\sim\multinormal[sB]\left(\mathbf{0},\boldsymbol{\Xi}\otimes \mathbf{I}\right)\]
with
\begin{align*}
\mathbf{I}\in\Rset^{B\times B},\qquad \boldsymbol{\Xi}=(\xi_{k\ell})\in\Rset^{s\times s},\qquad 
\xi_{k\ell}=\sigma^2 \lim_{\mylim} \mathbf{u}_{j_k 1}^\top \mathbf{u}_{j_\ell 1}.
\end{align*}

As the $B$ vectors $(\Ue[j_1]^1,\ldots,\Ue[j_s]^1),\ldots, (\Ue[j_1]^B,\ldots,\Ue[j_s]^B)$ converge to i.i.d.~random vectors, also the $B$ vectors $(|\Us[j_1]^1|,\ldots,|\Us[j_s]^1|),\ldots,(|\Us[j_1]^B|,\ldots,|\Us[j_s]^B|)$ converge to i.i.d.~random vectors. Then the combinations of their elements $\Us[\Sset]^1,\ldots,\Us[\Sset]^B$ given in \eqref{def:combTs_exact} converge to i.i.d.~random variables. As $g$ is increasing in each argument, high values of $\Us[\Sset]^1$ correspond to evidence against $H_\Sset$ and, from \citet{score} (see Lemma 1),
\[\lim_{\mylim}P\left(\Us[\Sset]^1 > \Us[\Sset]^{(\lceil (1-\alpha)B\rceil)}\right) = \frac{\lfloor\alpha B\rfloor}{B}\leq\alpha.\]
\end{proof}

\paragraph{Theorem \ref{T:multi_appr}}
\textit{The test that rejects $H_\Sset$ when $\Vs[\Sset]^1 > \Vs[\Sset]^{(\omega)}$ is asymptotically an $\alpha$-level test for any non-empty $\Sset\subseteq\Mset$.}

\begin{proof}
Proof of the theorem follows directly from the proof of Theorem \ref{T:multi_exact}. Fix any non-empty set $\Sset =\{j_1,\ldots,j_s\}\subseteq\Mset$, and assume that $H_\Sset$ is true. Then consider any couple of splits $q,r\in\{1,\ldots,Q\}$, any couple of variables $j,h\in\Sset$ and any couple of transformations $b,c\in\{1,\ldots,B\}$. Moreover, suppose that the variable selection procedure selects all active variables; by Assumption \ref{A:screening}, this is true at least asymptotically, so this assumption does not affect asymptotic results.

Consider the vector
\[\Vevect[\Sset]=(\Ve[j_1]^1,\ldots,\Ve[j_1]^B,\ldots,\Ve[j_s]^1,\ldots,\Ve[j_s]^B)^\top, \qquad \Ve^b =\frac{1}{\sqrt{n}}\ve^\top (\re^1+\ldots+\re^Q) \fmat (\re^1+\ldots+\re^Q) \yvec.\]
From \eqref{testar0} we have
\begin{align*}
\frac{1}{\sqrt{n}}\ve^\top\re^q \fmat \re^r \yvec =
\begin{cases}
\frac{1}{\sqrt{n}}\ve^\top\re^q \fmat \epsvec + o_P(1)\quad\text{if}\quad j\in \Aset^r\\
0\quad\text{otherwise}
\end{cases}
\end{align*}
and so
\[\Ve^b = a_j\Ue^{*b} + o_P(1),\qquad a_j=|\{q\,:\,j\in \Aset^q\}|\]
where
\[\Ue^{*b}=\frac{1}{\sqrt{n}}\ve^\top\reb \fmat\epsvec \]
is the same as defined in \eqref{def:testarq}. From the proof of Theorem \ref{T:multi_exact}, we have that the $B$ vectors $(\Ve[j_1]^1,\ldots,\Ve[j_s]^1),\ldots, (\Ve[j_1]^B,\ldots,\Ve[j_s]^B)$ converge to i.i.d.~random vectors, and so do the vectors $(|\Vs[j_1]^1|,\ldots,|\Vs[j_s]^1|),\ldots, (|\Vs[j_1]^B|,\ldots,|\Vs[j_s]^B|)$. The combinations of their elements $\Vs[\Sset]^1,\ldots,\Vs[\Sset]^B$ given in \eqref{def:combTs_appr} converge to i.i.d.~random variables. As high values of $\Vs[\Sset]^1$ correspond to evidence against $H_\Sset$, from \citet{score} (Lemma 1) we have
\[\lim_{\mylim}P\left(\Vs[\Sset]^1 > \Vs[\Sset]^{(\lceil (1-\alpha)B\rceil)}\right) = \frac{\lfloor\alpha B\rfloor}{B}\leq\alpha.\]
\end{proof}


\paragraph{Lemma \ref{L:complexity_exact}}
\textit{In the worst case, Algorithm \ref{algorithm:exact} (excluding the variable selection procedure) has computational complexity of order $n^4QB$, and memory usage of order $n^2Q$.}

\begin{proof}
Fix any $j\in\Mset$, and denote the number of splits where $j$ is selected with $a_j=|\{q\,:\,j\in \Aset^q\}|$. Recall that, for square matrices of size $n$, the computational complexity of multiplication, transposition and inversion is of order $n^3$. Hence computing $\re^q$ as in \eqref{def:splitresmat} for all splits that select $j$ requires $n^3 a_j$ operations. Computing $\Ue^b$ as in \eqref{def:Ue} for all transformations requires $n^3Ba_j$ operations.

Therefore the total complexity of the algorithm is order
\[n^3Ba_{\text{tot}},\qquad a_{\text{tot}}=\sum_{j\in\Mset}a_j. \]
In the worst case, where we select $n/2$ variables in each split, we have $a_{\text{tot}}=nQ/2$, and so the complexity is of order $n^4QB$.

Moreover, for each variable $j$ the algorithm needs to store $a_j$ square matrices of size $n$, with memory usage of order $n^2a_j$. In the worst case, $a_j=Q$, and so the memory usage is of order $n^2Q$.
\end{proof}

\paragraph{Lemma \ref{L:complexity_approx}}
\textit{In the worst case, Algorithm \ref{algorithm:approximate} (excluding the variable selection procedure) has computational complexity of order $n^4Q + n^3B$, and memory usage of order $n^2$.}

\begin{proof}
Analogously to the proof of Lemma \ref{L:complexity_exact}, fix any $j\in\Mset$, and denote the number of splits where $j$ is selected with $a_j=|\{q\,:\,j\in \Aset^q\}|$. Computing $\re^q$ as in \eqref{def:splitresmat} for all splits that select $j$ and $\reb$ requires $n^3 s_j$ and $n^2a_j$ operations, respectively. Computing $\Ve^b$ as in \eqref{def:Ve} for all transformations requires $n^3B$ operations.

Therefore the total complexity of the algorithm is order
\[n^3(B+a_{\text{tot}}),\qquad a_{\text{tot}}=\sum_{j\in\Mset}a_j. \]
In the worst case, where we select $n/2$ variables in each split, we have $a_{\text{tot}}=nQ/2$, and so the complexity is of order $n^4Q + n^3B$.

Moreover, for each variable $j$ the algorithm needs to store only 2 square matrices of size $n$,  $\re^q$ and $\reb$. Hence the memory usage is of order $n^2$.
\end{proof}

\end{document}